\documentclass[aps,prl,twocolumn,superscriptaddress,raggedbottom,showpacs,floatfix, longbibliography]{revtex4-2}
\usepackage{amsmath,amsfonts,amssymb,amsthm,graphics}
\usepackage[next]{inputenc}
\usepackage[colorlinks=true,citecolor=blue,linkcolor=blue]{hyperref}
\usepackage{bbm}
\usepackage{booktabs}
\usepackage{multirow}
\usepackage{hhline}
\usepackage{comment}
\usepackage{float}
\usepackage{latexsym}
\usepackage[dvipsnames]{xcolor}
\usepackage{graphicx}
\usepackage{epstopdf}
\usepackage{bm}
\usepackage{verbatim}
\usepackage{physics}
\usepackage{mathrsfs}
\usepackage{tikz}
\usepackage{chemfig}
\usepackage{pifont}
\usepackage{manfnt}
\usepackage{mathtools}
\usepackage{soul}

\usepgflibrary{patterns} 
\usepgflibrary[patterns] 
\usetikzlibrary{patterns} 
\usetikzlibrary[patterns]
\usetikzlibrary{shapes,shadows,arrows,positioning,graphs}

\newcommand{\hexagon}{\mathord{\raisebox{-1pt}{\tikz{\node[draw,scale=.65,regular polygon, regular polygon sides=6,fill=none](){};}}}}

\newcommand{\tetrahedron}{
 \raisebox{-1pt}{\includegraphics[height=2ex]{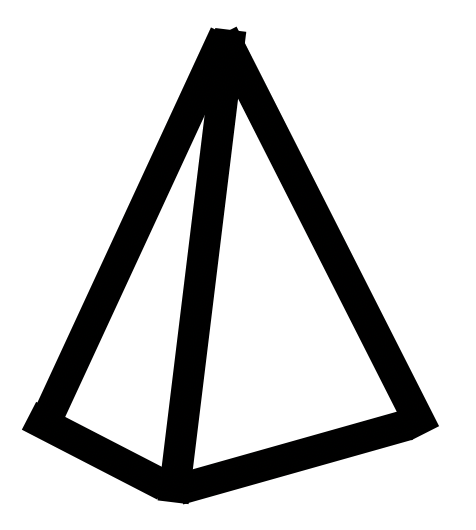}}}

\begin{document}

\title{Interplay of symmetry breaking and deconfinement in 3D quantum vertex models}
\author{Shankar Balasubramanian}
\affiliation{Center for Theoretical Physics, Massachusetts Institute of Technology, Cambridge, MA 02139, USA}
\author{Daniel Bulmash}
\affiliation{Joint Quantum Institute and Condensed Matter Theory Center, Department of Physics, University of Maryland, College Park, Maryland 20742-4111, USA}
\author{Victor Galitski}
\affiliation{Department of Physics, University of Maryland, College Park, Maryland 20742-4111, USA}
\author{Ashvin Vishwanath}
\affiliation{Department of Physics, Harvard University, Cambridge, MA 02138, USA}

\begin{abstract}
We construct a broad class of frustration-free \emph{quantum vertex models} in 3+1D whose ground states are weighted superpositions of classical 3D vertex model configurations.  Our results are illustrated for diamond, cubic, and BCC lattices, but hold in general for 3D lattices with even coordination number. The corresponding classical vertex models have a $\mathbb{Z}_2$ gauge constraint enriched with a $\mathbb{Z}_2$ global symmetry.  We study the interplay between these symmetries by exploiting exact wavefunction dualities and effective field theories.  We find an exact gapless point which by duality is related to the Rokhsar-Kivelson (RK) point of $U(1)$ spin liquids.  At this point, both the symmetry breaking and deconfinement order parameters exhibit long range order.  The gapless point is additionally a self-dual point of a second duality that maps the $\mathbb{Z}_2$ deconfined and $\mathbb{Z}_2$ symmetry-broken phases to one another.  For the BCC lattice vertex model, we find that gapless point is proximate to an unusual intermediate phase where symmetry breaking and deconfinement coexist.
\end{abstract}

\maketitle

\noindent \emph{Introduction}--- 
Quantum spin liquids represent zero temperature phases of matter whose essential characteristics lie beyond the Landau order parameter paradigm \cite{wenbook}. Quantum dimer models  represent a family of simplified models \cite{Fradkin,RK,RSPRB90,RC89,FA} where novel properties of quantum spin liquids such as fractionalized excitations, emergent gauge bosons and ground state degeneracy can be readily accessed. Analytic progress is aided by the fact that at specially tuned Rokhsar-Kivelson (RK) points \cite{RK}, correlation functions in the ground state of quantum dimer models match correlation functions in classical dimer models \cite{RK, MoessSon, FMS, Henley, MSP}, which can be exactly computed in two dimensions \cite{Kasteleyn, TemperleyFisher}. Due to the dearth of exactly solvable statistical mechanics models in three dimensions, less is understood about the general phase structure of 3+1D quantum dimer models, although exotic phases such as $U(1)$ spin liquids are known to exist \cite{HKMS, Hermele, MS}. 

In this paper, we study a family of 3+1D quantum dimer models enriched with a global $\mathbb{Z}_2$ dimer flip symmetry. At suitably defined RK points, these models have ground states that are weighted superpositions of configurations in a 3D classical vertex model, a statistical mechanics model of interacting loops.  We call these \emph{quantum vertex models}; prior work on such models has focused on the 2D square lattice \cite{AFF, castel}, with recent work generalizing to other 2D lattices \cite{Balasubramanian}. The 3+1D quantum vertex models exhibit a quantum phase transition between a deconfined phase and a phase which spontaneously breaks the $\mathbb{Z}_2$ symmetry -- this transition can be mapped to a thermodynamic phase transition in the 3D classical vertex models.  The exact location of the transition point can be determined with the help of \emph{two} different dimension-independent self dualities found in classical vertex models. In 3+1D, the gapless transition point has the unusual property that local dimer order and deconfinement coexist.  We also construct a quantum vertex model where the gapless point is proximate to a phase where dimer and deconfinement order parameters acquire nonzero expectation values.  This proximate phase bears resemblance to fragmented Coulomb phases~\cite{savarybalents, powell,brooksbartlett, fragmentationreview, raban} but is found in a qualitatively different setting.  We use effective field theory methods to study the phase structure near this multicritical point and to analyze deformations of the quantum vertex model away from the RK manifold.

\noindent \emph{Quantum vertex models}--- We construct a Hamiltonian defined on a regular bipartite lattice in 3D with even coordination number.  We will focus on three cases, where the bipartite lattice is (i) a diamond lattice, dual to the pyrochlore lattice of corner sharing tetrahedra, (ii) a cubic lattice, dual to a lattice of corner sharing octahedra, and (iii) a BCC lattice, dual to a lattice of corner sharing cubes.  However, our results generically hold for \emph{any} even-coordinated lattice in 3D.

We start by briefly defining a classical vertex model, whose degrees of freedom are dimers on the links of the lattice.  The classical Boltzmann weight $\mathcal{W}(C)$ of a configuration $C$ of dimers is the product of vertex weights $W_{V_p}(n_C)$ at each site $p$, where $n_C$ is the number of dimers touching $p$ in configuration $C$ and $V_p$ is the coordination number: $\mathcal{W}(C) = \prod_{p} W_{V_p}(n_C)$.  The partition function is then $\mathcal{Z} = \sum_C \mathcal{W}(C)$.  Motivated by this construction, we define an (unnormalized) RK ground state wavefunction $\ket{\text{GS}} = \sum_{C} \sqrt{\mathcal{W}(C)} \ket{C}$, whose norm is the partition function of the classical vertex model.

We now discuss the choice of vertex weights.  We explicitly impose a $\mathbb{Z}_2$ symmetry of exchanging dimers and empty links in the vertex weights: for the 8-vertex model on a diamond lattice
\begin{equation}
    W_4(0) = W_4(4) = u, \hspace{0.5cm} W_4(2) = 1,
\end{equation}
for the 32-vertex model on a cubic lattice
\begin{equation}
    W_6(0) = W_6(6) = u, \hspace{0.5cm} W_6(2) = W_6(4) = 1,
\end{equation}
and for the 128-vertex model on the BCC lattice
\begin{equation}
    W_8(0) = W_8(8) = u, \hspace{0.25cm} W_8(2) = W_8(6) = v, \hspace{0.25cm} W_8(4) = 1.
\end{equation}
For the diamond and BCC lattices, $u = v = 0$ corresponds to an ice rule constraint where the number of dimers equals the number of empty links at each site.  On the cubic lattice, the ice rule constraint requires three dimers per site \cite{MS}.  These points have an emergent $U(1)$ gauge structure, and in the diamond and cubic lattices correspond to RK points of known $U(1)$ spin liquid phases \cite{Hermele, MS}.  Using an exact duality, we show that these points are mapped to $u = 3$ on the diamond lattice, $u = 5$ on the cubic lattice, and $u = 35/3, v = 5/3$ on the BCC lattice.  These gapless points will be a large focus of this paper.

We may also construct a frustration-free Hamiltonian whose ground state is the RK wavefunction $\ket{\text{GS}} = \sum_{C} \sqrt{\mathcal{W}(C)} \ket{C}$, and therefore shares the same RK phase diagram as the classical vertex model.  This construction is standard and covered in the Supplementary Material (see ~\cite{si} and Refs.~\cite{Balasubramanian,peskin1980critical,Kogut} therein).  Define an operator $P$ that projects onto an even dimer constraint at each site.  Schematically, the Hamiltonian may be written as $H = P + H_A + r H_B$, where
\begin{align}
    H_A &= \sum_{C_{\ell}, \ell} \frac{1}{\omega_{u,v}(C_\ell)} \ketbra{C_{\ell}}{C_{\ell}} + \omega_{u,v}(C_\ell) \ketbra{\overline{C}_{\ell}}{\overline{C}_{\ell}} \label{eq:fullham}\\
    H_B &= -\sum_{C_{\ell}, \ell} \ketbra{C_{\ell}}{\overline{C}_{\ell}} + \ketbra{\overline{C}_{\ell}}{C_{\ell}} \label{eq:fullham2}
\end{align}
with $\omega_{u,v}$ a certain scalar function of $u$ and $v$, $\ell$ corresponding to a minimal length loop on the lattice (squares on the cubic and BCC lattices, and hexagons in the diamond lattice), $C_{\ell}$ a configuration of spins on $\ell$, and $\overline{C}_{\ell}$ the opposite configuration of spins.  The RK wavefunction is a ground state for $r = 1$.  As in the classical vertex model, this Hamiltonian has a global $\mathbb{Z}_2$ spin-flip symmetry $\prod_i X_i$ interchanging dimers and empty links.

\begin{figure}
    \centering
    \includegraphics[scale=0.35]{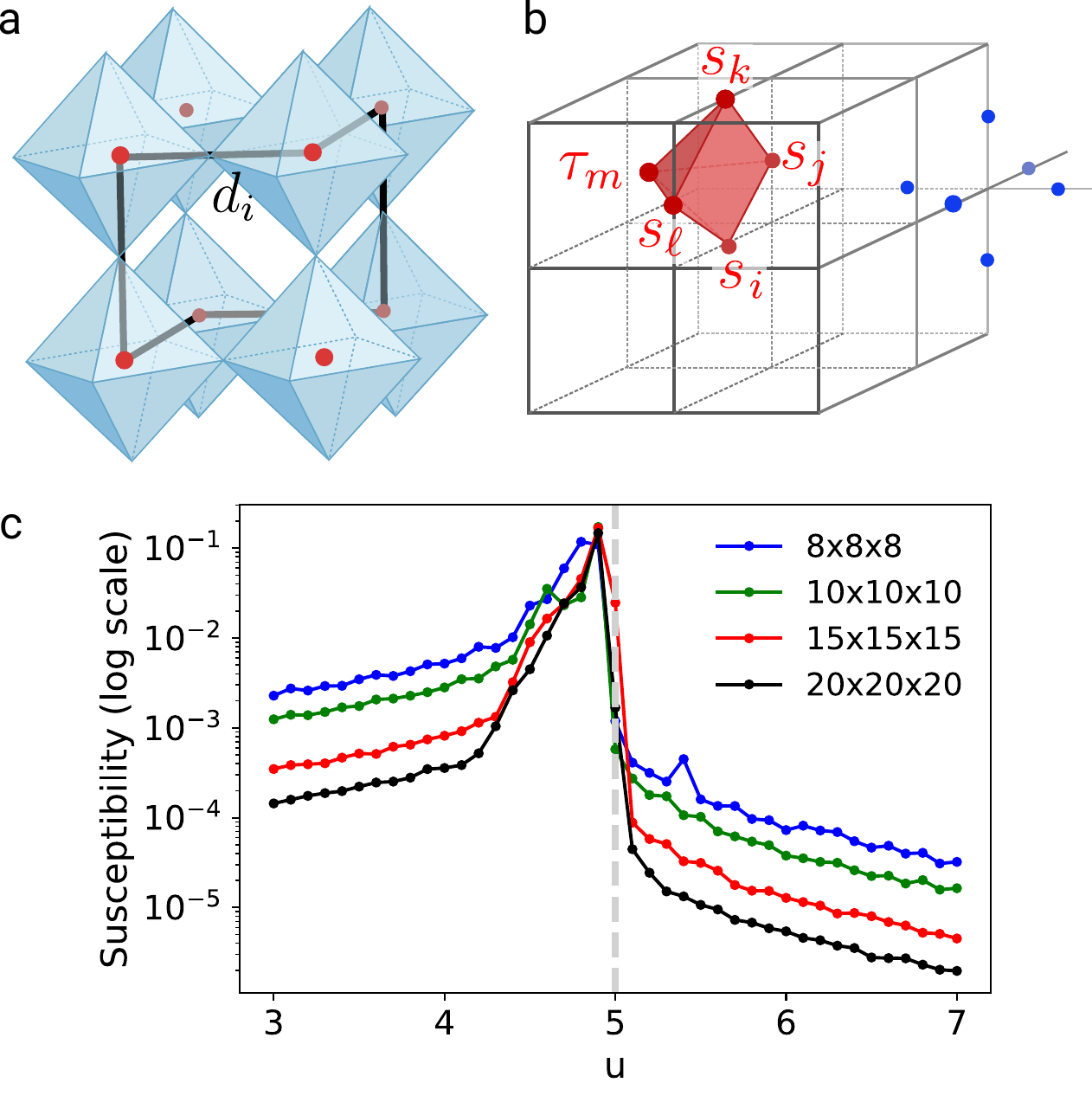}
    \caption{The dual lattice of the octahedral lattice is a cubic lattice, where the vertex model is defined (a).  The spin model derived in the Appendix is shown in (b), with gauge symmetries indicated by the blue dots and the 5-spin interaction indicated by the red pyramids.  Monte Carlo simulations of the dimer susceptibility suggest a transition at $u = 5$.}
    \label{fig:octahedra}
\end{figure}

\noindent \emph{Dualities and phase diagrams}--- Classical statistical mechanics in 3D is known to admit very few exact solutions \cite{Wu, zam, baxterbazhanov}; however, conveniently, vertex models possess a self-duality which is a special case of Wegner's duality \cite{Wegner, WuWu, Runnels, Balasubramanian} and holds in all dimensions.  This duality exactly relates the partition functions of classical vertex models for vertex weights $(u,v)$ and $(u^*,v^*)$.  One can show that for an 8-vertex model on a diamond lattice, Wegner's duality gives $u^* = (3+u)/(u-1)$, which indicates that $u = 3$ is a self dual point \cite{Sutherland1, Sutherland2, Thibaudier, Villain}.  For the 32-vertex model on a cubic lattice, Wegner's duality gives $u^* = (15+u)/(u-1)$, so that $u = 5$ is the self dual point.  These coincide with the gapless points previously discussed.  For the BCC lattice, Wegner's duality gives $u^* = (35+u+28v)/(3+u-4v)$ and $v^* = (-5+u+4v)/(3+u-4v)$, yielding the self-dual line $u - 4v = 5$ which contains the gapless point $u = 35/3, v = 5/3$.

On the diamond and BCC lattices, there is a second self-duality, which we call the decorated Wegner duality.  It can be shown that the partition function formally obeys $\mathcal{Z}(u) = \mathcal{Z}(-u)$ on the diamond lattice and $\mathcal{Z}(u,v) = \mathcal{Z}(u,-v)$ on the BCC lattice. Composing this identity with Wegner's duality gives $\mathcal{Z}(\mathsf{u}^*, \mathsf{v}^*) = \mathcal{Z}(u,v)$ where $\mathsf{u}^* = (3-u)/(u+1)$ for the diamond lattice, and $\mathsf{u}^* = (35+u-28v)/(3+u+4v)$ and $\mathsf{v}^* = (5-u+4v)/(3+u+4v)$ for the BCC lattice.  For these lattices, this duality maps the ice rule point $u = v = 0$ to the gapless point.

The quantum vertex models feature additional special points.  At the point $u = v = 1$, the parent Hamiltonian becomes the 3D toric code on the respective dual lattice \cite{kitaev2003fault}, and thus the phase diagram must include a gapped phase with $\mathbb{Z}_2$ topological order.  At $u = \infty$, one finds a phase that breaks the global $\mathbb{Z}_2$ symmetry.   The gapless point indicates a phase transition between the deconfined and symmetry broken phases; for all three lattices we performed Monte Carlo simulations to verify this (see Figures~\ref{fig:octahedra}[c], \ref{fig:pyrochlore}[d], and \ref{fig:BCCnumerics}).  Other ground state transitions out of the toric code phase were numerically studied in Refs.~\cite{Trebst, kitaev2, reiss2019quantum} and thermal transitions were studied in Refs.~\cite{PhysRevB.76.184442, PhysRevB.78.155120}. 
\begin{figure}
    \centering
    \includegraphics[scale=0.45]{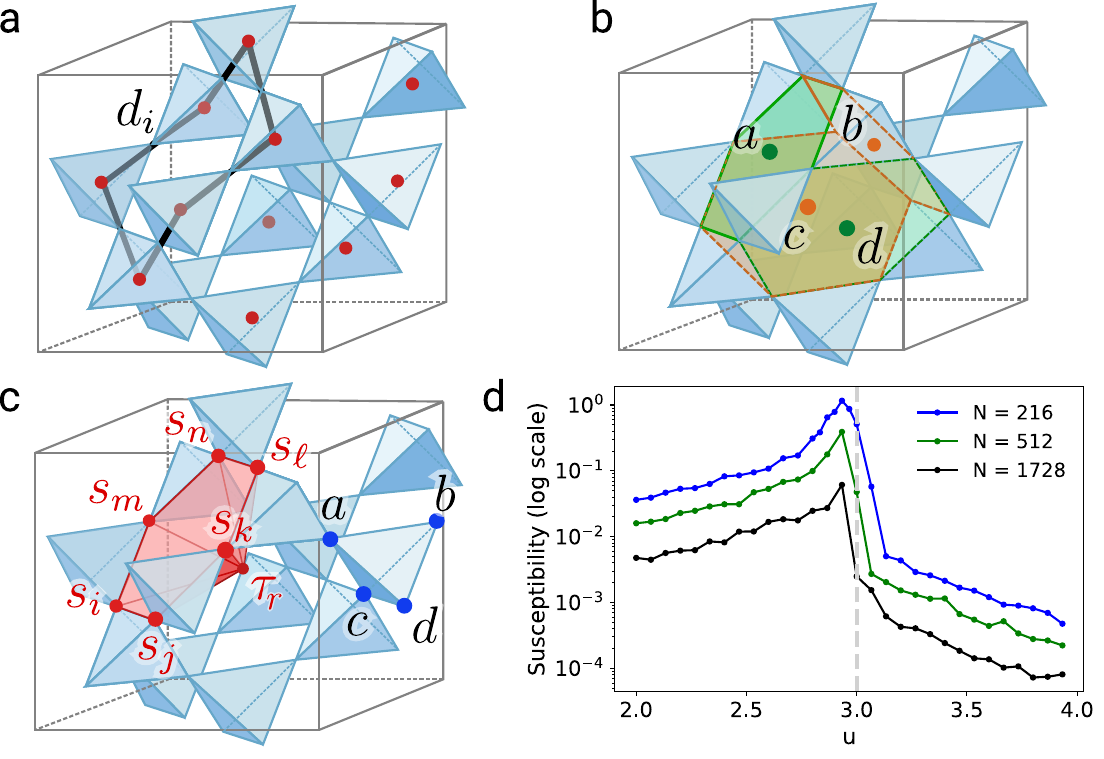}
    \caption{The pyrochlore lattice is shown in (a), with dimers on the diamond lattice labelled.  Panels (b) and (c) are relevant for the mapping to a classical spin model, which is derived in the Appendix.  The spins $s_i$ are placed on the centers of the hexagons (b) forming a dual pyrochlore lattice shown in (c).  On the dual pyrochlore lattice, the gauge symmetry is on the tetrahedra and the 7-spin interaction is labelled.  Monte Carlo simulations of the dimer susceptibility show a phase transition around $u = 3$ in (d) and $N$ denotes the number of sites in the simulation.}
    \label{fig:pyrochlore}
\end{figure}

\noindent \emph{Effective field theories}--- We now proceed to characterizing the gapless points ($u = 5$ for cubic, $u = 3$ for diamond, and $(u,v) = (35/3, 5/3)$ for BCC) by deriving an effective field theory.  To do so, we observe that it is possible to introduce a new formulation of the vertex model in terms of height variables via the following exact rewriting of the partition function at the self dual point:
\begin{equation}
    \mathcal{Z}_0 = \int_0^{2\pi} d^n \theta \prod_{\langle p, q \rangle} \cos\left(\theta_p - \theta_q\right),
\end{equation}
where $p$ and $q$ are sites on the diamond, cubic, or BCC lattices.  Interpreting $\theta_p$ as height fields, in the long wavelength limit, we assume that fluctuations of this field are small and can postulate a Euclidean field theory
\begin{equation}
    \mathcal{Z}_0 = \int \mathcal{D}\theta(x)\, \exp\left(-\frac{K}{2} \int d^3 x\, (\nabla \theta)^2 + \cdots\right),
\end{equation}
which has an global $U(1)$ symmetry due to the shift invariance of $\theta$.  A different height field representation was previously proposed for the 2D 8-vertex model on a square lattice \cite{Knops1, Knops2}.  To understand this emergent $U(1)$ symmetry, recall that the decorated Wegner duality maps an ice rule constrained model to the gapless point. In the ice rule limit, one can construct a vector potential and formulate an effective action equivalent to $U(1)$ gauge theory in $2+1$D \cite{HKMS}.  This action has a hidden $U(1)_{\text{top}}$ symmetry corresponding to the monopole charge (where the operator that creates a monopole in the $U(1)$ gauge theory is $\mathcal{M}_{\bm{x}} = e^{i \theta(\bm{x})}$).  The decorated Wegner duality thus maps a $U(1)$ gauge theory in $2+1$D to a ``dual photon'', where the $U(1)_{\text{top}}$ symmetry is manifested as a global shift symmetry; this duality both extends over a large parameter regime and is exact on the lattice. Electric field correlations are equivalent to dimer correlations, which are dipolar in the ice rule limit: $\langle s_i(\bm{x})s_j(\bm{y}) \rangle \sim \frac{1}{r^5}(3r_ir_j - r^2 \delta_{ij})$, where $\bm{r} = \bm{y} - \bm{x}$.  This is consistent with the effective field theory as under the decorated Wegner duality, the correlation function maps to $\langle s_i(\bm{x})s_j(\bm{y}) \rangle = \langle \tan\left(\theta_{\bm{x}} - \theta_{\bm{x} + \bm{e_i}}\right) \tan\left(\theta_{\bm{y}} - \theta_{\bm{y} + \bm{e_j}}\right)\rangle \approx \langle \partial_i \theta(\bm{x}) \partial_j \theta(\bm{y})\rangle$, which also has a dipolar form.

Next, we compute the spontaneous dimer density near the gapless point, $\langle d_i(\bm{x}) d_j(\bm{y}) \rangle$, which can be shown to be $\left\langle \frac{\cos\left(\theta_{\bm{x}} + \theta_{\bm{x} + \bm{e_i}}\right)}{\cos\left(\theta_{\bm{x}} - \theta_{\bm{x} + \bm{e_i}}\right)} \frac{\cos\left(\theta_{\bm{y}} + \theta_{\bm{y} + \bm{e_i}}\right)}{\cos\left(\theta_{\bm{y}} - \theta_{\bm{y} + \bm{e_i}}\right)}\right\rangle$
 and maps onto $\langle \cos (2\theta(\bm{x})) \cos (2\theta(\bm{y})) \rangle$ in the long-wavelength limit.  For a compact boson in three dimensions, this operator exhibits long range order.  We may also consider the operator $\langle \sin (2\theta(\bm{x})) \sin (2\theta(\bm{y})) \rangle$, which on the lattice corresponds to applying two test charges at $\bm{x}$ and $\bm{y}$ and computing the ratio of partition functions with and without the charges, thus serving as a diagnostic for deconfinement (see Supplemental Material~\cite{si}). This operator also exhibits long range order, so at the gapless point local order and deconfinement coexist.

Next, we discuss what occurs when one perturbs about the gapless point, but still within $(u,v)$ space.  We utilize the following exact rewriting of the partition function
\begin{equation}
    \mathcal{Z}_{\alpha} = \int_0^{2\pi} d^n\theta \prod_{\langle p,q \rangle} \left(f(\theta_p)f(\theta_q) + g(\theta_p)g(\theta_q)\right),
\end{equation}
where $f(\theta) = \cos \theta + \alpha \cos(3\theta) + \beta \cos(7\theta)$ and $g(\theta) = \sin \theta - \alpha \sin(3\theta) - \beta \sin(7\theta)$.  An explicit relation between $(u,v)$ and $(\alpha, \beta)$ is presented in the Supplemental Material~\cite{si}.  For the diamond and cubic lattices, $\beta = 0$, and is only nonzero for the BCC lattice. Conveniently, under Wegner's duality, $(\alpha, \beta) \to (-\alpha, \beta)$, which as an operation on the fields is equivalent to $\theta \to \theta + \pi/4$. Setting $\beta = 0$ for simplicity, and using the explicit forms for $f$ and $g$, the partition function becomes a sine-Gordon model in the long wavelength limit, to leading order in $\alpha$:
\begin{equation}
    \mathcal{Z}_{\alpha} = \int \mathcal{D}\theta(x)\, \exp\left(-\frac{K}{2}\int d^3 x\, (\nabla \theta)^2 - \frac{4 \alpha}{K} \cos (4 \theta) + \cdots\right)
\end{equation}
At $\alpha=0$ the $U(1)$ symmetry implies that all symmetry breaking cosine terms vanish.  The $U(1)$ symmetry breaking terms are relevant, consistent with the gapless point indicating a transition.  For small positive $\alpha$, $|\langle \cos (2 \theta) \rangle|$ is pinned to 1, indicating a $\mathbb{Z}_2$ broken phase, and for small negative $\alpha$, $\langle \cos (2 \theta) \rangle$ is pinned to 0, indicating a $\mathbb{Z}_2$ unbroken phase.  The deconfinement operator $|\langle \sin (2\theta) \rangle|$ is $0$ when $\alpha$ is positive and $1$ when $\alpha$ is negative, consistent with it being a diagnostic of topological order.  Under Wegner's duality, $\cos (2\theta)$ and $\sin (2\theta)$ are exchanged at the gapless point.  

\begin{figure}
    \centering
    \includegraphics[scale=0.55]{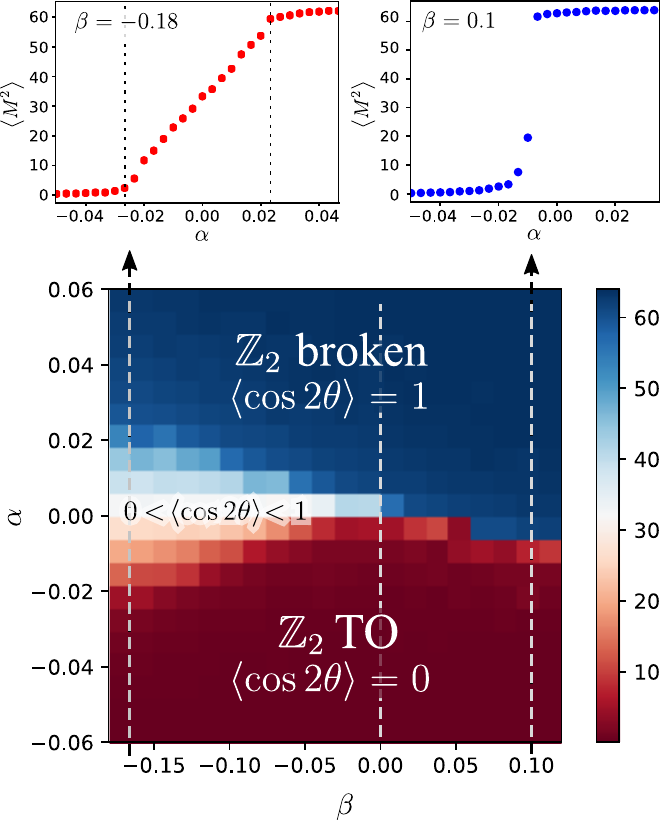}
    \caption{Monte Carlo results for average value of the squared magnetization of an $L \times L \times L$ BCC lattice vertex model.  The bottom panel ($L = 8$) this presents evidence of the existence of the intermediate phase, proximate to the gapless point at $(\alpha, \beta) = (0,0)$.  Vertical cuts are shown in the two top panels (simulations here are for $L = 12$) -- in the left plot, two sharp transitions are seen and the duality maps these transition points to one another, while in the right plot, a single sharp transition occurs.}
    \label{fig:BCCnumerics}
\end{figure}

For nonzero $\beta$ on the BCC lattice, the effective action becomes
\begin{equation}
    \mathcal{S}[\alpha, \beta] = \frac{K}{2}\int d^3 x\, (\nabla \theta)^2 - \frac{4 \alpha}{K} \cos (4 \theta) - \frac{4 \beta}{K} \cos (8 \theta) + \cdots
\end{equation}
consistent with $\beta$ coupling to a duality-even operator.  Since $\cos(4\theta)$ and $\cos(8\theta)$ are both relevant, we naively mean field theory to the effective potential $V(\theta) = \alpha \cos (4\theta) + \beta \cos (8\theta)$, which yields a first order line $\alpha = 0, \beta > 0$ between a $\mathbb{Z}_2$ broken and deconfined phase, and two second order lines in the 3D Ising universality class $\alpha \propto \pm \beta$ for $\beta < 0$. The region between these critical lines is a phase where both the dimer order parameter and the deconfinement parameter have nonzero and continuously varying expectation values; evidence for this exotic coexistence phase is seen in Monte Carlo simulations of the dimer density in the $(\alpha, \beta)$ plane, see Fig.~\ref{fig:BCCnumerics}.

On general grounds, one could phenomenologically arrive at the effective field theory by writing down terms consistent with invariance under the $\mathbb{Z}_2$ symmetry $\cos(2\theta) \to -\cos(2\theta)$:
\begin{equation}
    \mathcal{S} = \int d^3 x\, \left(\frac{K}{2}(\nabla \theta)^2 + \cdots + \sum_n c_n \cos (4 n \theta)\right).
\end{equation}
Since the local dimer density at each vertex is bounded between $-V_p$ and $+V_p$ and the operator $\cos (4 n \theta)$ corresponds to changing the relative weight of vertex configurations with $k$ and $k \pm 4n$ dimers at a fixed site, we require $4n \leq V_p$, consistent with the proposed action for $V_p = 4, 6, 8$.  The fact that $\cos (4 n \theta)$ is duality even for $n \in 8\mathbb{Z}$ and duality odd for $n \in 8\mathbb{Z} + 4$ remains true on the lattice.  Analogously to the BCC lattice, we anticipate the possibility of more interesting coexistence phases for vertex models where $V_p > 8$.

\begin{figure}
    \centering
    \includegraphics[scale=0.2]{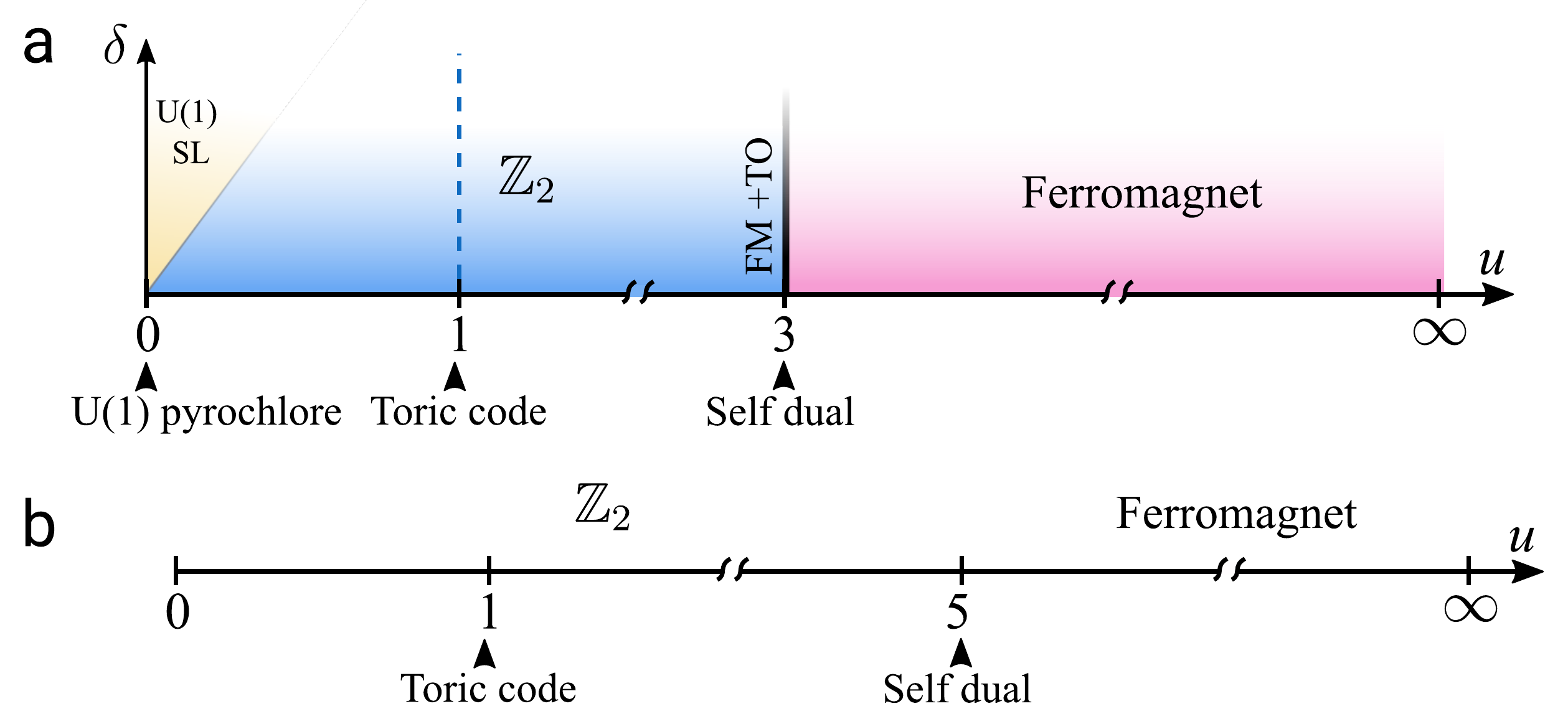}
    \caption{Panel (a) $u$-axis (where $u$ is the vertex weight) and panel (b) show the phase diagram of the frustration free models on the diamond and cubic lattices respectively.  The $\delta$-axis in panel (a) corresponds to perturbing the diamond lattice model away from the RK manifold.}
    \label{fig:phaseDiagram}
\end{figure}

Finally, we discuss the effective field theory description away from the RK manifold.  Consider extending the microscopic Hamiltonian in Eqns.~\ref{eq:fullham} and \ref{eq:fullham2} beyond the RK point, via some generic isotropic perturbation of strength $\delta$ (as an example, we may call $\delta \sim r-1$).  As the effective field theory describing the RK wavefunction at the gapless point is a free boson, the effective field theory description for the \emph{Hamiltonian} is a quantum Lifshitz model \cite{henleyheight, FHMOS, VBS, AFF, Fradkin}, with the imaginary time action
\begin{equation}
    \mathcal{S} = \int d^3x\, d\tau \, \left(\frac{1}{2} (\partial_{\tau} \theta)^2 + \frac{\kappa}{2} (\nabla^2 \theta)^2 + \delta (\nabla \theta)^2\right) + \mathcal{S}_{\text{mon}},
\end{equation}
where $\mathcal{S}_{\text{mon}}$ accounts for monopole events: $\mathcal{S}_{\text{mon}} = \int d^3x\, d\tau \, (g_4 \cos(4\theta) + g_8 \cos(8\theta))$ \cite{Fradkin} (we expect $g_4$ and $g_8$ to play the roles of $\alpha$ and $\beta$).  The RK manifold corresponds to $\delta = 0$, and the $(\nabla^2 \theta)^2$ term becomes irrelevant at $\delta > 0$, resulting in a gapless line with dynamical critical exponent $z_{\text{dyn}}=1$ between the $\mathbb{Z}_2$ deconfined and $\mathbb{Z}_2$ broken phases.  On the pyrochlore and BCC lattices the spin ice point $u = v = 0$ is expected to broaden into a $U(1)$ spin liquid phase away from the RK manifold and will always be separated from the $\mathbb{Z}_2$ broken phase by a $\mathbb{Z}_2$ deconfined phase.  This follows because for the choice $\delta \sim r-1$ the Hamiltonian at $u = 1$ always exhibits toric code topological order regardless of the value of $r$.

Furthermore, as we move along the $z_{\text{dyn}}=1$ line (where $g_4 = g_8 = 0$),  the equal-time correlators $\langle \cos (2\theta(\bm{x})) \cos (2\theta(\bm{y})) \rangle$ and $\langle \sin (2\theta(\bm{x})) \sin (2\theta(\bm{y})) \rangle$ exhibit long range order, indicating that coexistence still occurs.  Studying the phase structure for $\delta > 0$ when the monopole terms are turned on, especially the stability of the coexistence phase on the BCC lattice, warrants further analytical and numerical investigations.

{\em Acknowledgements --} A.V. and V. G. were supported by the Simons Collaboration on Ultra Quantum Matter, a grant from the Simons Foundation (651440, A.V.).  S.B. was supported by the National Science Foundation Graduate Research Fellowship under Grant No. 1745302. V.G. was also supported by NSF DMR-2037158 and US-ARO Contract No.W911NF1310172. D.B. was supported by JQI-PFC-UMD.

\noindent \emph{Appendix on equivalent spin models}--- In this Appendix we provide a different perspective on the phase structure of the 3D classical vertex models by a mapping to a classical spin model.  We first illustrate this mapping for the cubic lattice vertex model, where the vertex weights are $W_6(0)=W_6(6)=u$ and $W_6(2)=W_6(4)=1$. We first construct a dual cubic lattice, whose sites are located in the cube centers of the direct lattice.  Dimers on the direct lattice therefore pierce faces of the dual lattice.  We place new spins on the edges of this dual lattice so that each dimer on the direct lattice is surrounded by four spins $s_1,\cdots,s_4$ on the dual lattice.  The dimer variable is mapped to spins as $d = s_1s_2s_3s_4$; this automatically respects the even dimer constraint due to $\prod_{\square \in \mbox{\mancube}} s_i s_j s_k s_\ell = \prod_{- \in \mbox{\mancube}} s_i^2 = 1$.  The Boltzmann weight enforcing the even dimer constraint at a given vertex with vertex weights $W_6(n)$ is therefore
\begin{equation}
    W_{\mbox{\mancube}} (\vec{s}) = \cosh \left(\mathcal{J} \sum_{\square \in \mbox{\mancube}} s_i s_j s_k s_\ell \right),
\end{equation}
where $\frac{\cosh 6\mathcal{J}}{\cosh 2\mathcal{J}} = u$.  The full partition function is therefore
\begin{equation}
    \mathcal{Z}_{\text{spin}} = \sum_{\vec{s}} \prod_{\mbox{\mancube}} W_{\mbox{\mancube}} (\vec{s}) = \sum_{\vec{s}, \vec{\tau}} \exp\left(\mathcal{J} \sum_p s_i s_j s_k s_\ell \tau_m \right),
\end{equation}
where in the second equality we converted the $\cosh$ to an exponential by adding a new `ghost' spin $\tau$ at the center of each cube on the dual lattice; the subscript $p$ denotes a pyramid formed from four spins identifying a dimer and a ghost spin (see Figure \ref{fig:octahedra}[b]).

First, the standard  $\mathbb{Z}_2$ gauge symmetry $P = \prod_{i \in v} X_i = X_1X_2X_3X_4X_5X_6$ is easily seen, where $X_i$ is the spin flip operator acting on $s_i$, and $i \in v$ denotes spins $s_i$ on the edges connected to site $v$, see Fig.~\ref{fig:octahedra}[b] (shown as blue dots). Moreover, the gauge theory is enriched with a global $\mathbb{Z}_2$ symmetry corresponding to the operator mapping $d_i \to -d_i$ and $\tau_i \to -\tau_i$ (the global transformation $d_i \to -d_i$ requires flipping some subset of the $s$ spins, see Supplementary Material~\cite{si}).  As a result, an unusual aspect of this gauge theory is that the scaling of the Wilson loop $W_C = \left\langle\prod_{i \in C} s_i\right\rangle$ exhibits what we call an `even-odd' behavior. At large $\mathcal{J}$ the Wilson loops obey a perimeter law behavior, a consequence of simultaneous deconfinement and breaking of the global $\mathbb{Z}_2$ symmetry.  When $\mathcal{J}$ is small, Wilson loops enclosing an $\emph{even}$ number of plaquettes obeys an area law, while Wilson loops enclosing an $\emph{odd}$ number of plaquettes are zero.  This follows because odd Wilson loops are charged under the global symmetry while even Wilson loops are not.

On the diamond lattice, one can construct an analogous spin model.  The dimers live on the sites of a pyrochlore lattice, and  we assign a spin $s_{i}$ to the center of each hexagon of the pyrochlore lattice.  The spins form a dual pyrochlore lattice as illustrated in Figure \ref{fig:pyrochlore}[b], with dimers on the direct lattice piercing hexagons on the dual lattice.  The dimer variables are $d_i = \prod_{j \in \widetilde{\hexagon}_i} s_{j}$ where $\widetilde{\hexagon}_i$ denotes the hexagon on the dual lattice pierced by dimer $i$ on the direct lattice.  A gauge transformation corresponds to flipping the 4 spins on tetrahedra of the dual pyrochlore lattice via $P = \prod_{i \in \widetilde{\tetrahedron}} X_{i}$.

We introduce a ghost spin $\tau$ at the locations indicated in Fig.~\ref{fig:octahedra}[c], so that the partition function is
\begin{equation}
    \mathcal{Z}_{\text{spin}} = \sum_{\vec{s}, \vec{\tau}} \exp\left(\mathcal{J} \sum_p s_i s_j s_k s_{\ell} s_m s_n \tau_r \right),
\end{equation}
with the 7-spin interaction on pyramid $p$ labelled in Figure \ref{fig:pyrochlore}[c] and $\cosh 4\mathcal{J} = u$.  The Wilson loops of this classical gauge theory also exhibit an even-odd effect.  

On the BCC lattice, two different effective spin models are provided in the Supplementary Material~\cite{si}: along with a global $\mathbb{Z}_2$ symmetry, one of the models has a $\mathbb{Z}_2$ gauge structure, while the other model curiously has a subsystem symmetry.  The coexistence phase in the BCC lattice vertex model will also correspond to a coexistence phase in these spin models. 

In the original quantum vertex model, an odd Wilson loop enclosing a single plaquette corresponds to a dimer variable, which probes the $\mathbb{Z}_2$ symmetry breaking transition.  An even Wilson loop of spins corresponds to a 't Hooft loop of dimers $W_{\partial S} = \left\langle \prod_{i \in S} d_i \right \rangle$, where $\partial S$ is the boundary of open surface $S$.  In addition, one has a dual Wilson loop order parameter which is a product of $X_i$ around a loop of links \cite{AFF}: $\widetilde{W}_{C} = \left\langle \prod_{i \in C} X_i \right \rangle$. 

\bibliography{cit}
\onecolumngrid
	\begin{center}
		
        \bigskip 
        \bigskip 
\textbf{\large Supplementary material for:\\ ``Interplay of symmetry breaking and deconfinement in 3D quantum vertex models''}\\[.3cm]		
		Shankar Balasubramanian$^{1}$, Daniel Bulmash$^{2}$, Victor Galitski$^{3}$, Ashvin Vishwanath$^{4}$ \\[.1cm]
		{\itshape ${}^1$Center for Theoretical Physics, Massachusetts Institute of Technology, Cambridge, MA 02139, USA\\
			${}^2$Joint Quantum Institute and Condensed Matter Theory Center, Department of Physics, University of Maryland, College Park, Maryland 20742-4111, USA \\
            ${}^3$Department of Physics, University of Maryland, College Park, Maryland 20742-4111, USA \\
            ${}^4$Department of Physics, Harvard University, Cambridge, MA 02138, USA }
	\end{center}

\title{Supplementary Material for: ``Interplay of symmetry breaking and deconfinement in 3D quantum vertex models''}
\author{Shankar Balasubramanian}
\affiliation{Center for Theoretical Physics, Massachusetts Institute of Technology, Cambridge, MA 02139, USA}
\author{Daniel Bulmash}
\affiliation{Joint Quantum Institute and Condensed Matter Theory Center, Department of Physics, University of Maryland, College Park, Maryland 20742-4111, USA}
\author{Victor Galitski}
\affiliation{Department of Physics, University of Maryland, College Park, Maryland 20742-4111, USA}
\author{Ashvin Vishwanath}
\affiliation{Department of Physics, Harvard University, Cambridge, MA 02138, USA}

\maketitle

\onecolumngrid

\section{Wegner's duality and Decorated Wegner's duality}

In this section, we review some known results and provide some new results regarding duality mappings for vertex models, with an emphasis that these results are \emph{independent of the spatial dimension}.  Part of this appendix serves as a summary of \cite{Balasubramanian}.  We first will derive Wegner's duality for vertex models on a generic lattice.  To start, consider a vertex model on the dual lattice $G$.  We define spin variables $s_{pq}$ located on the edges of $G$, and we will assume that the vertex weights are only a function of the total number of bonds $b$ coming out of a site and are labelled as $W(b)$.  If $s_{pq} = 1$, then $\langle p,q \rangle$ forms an occupied bond on the vertex model.  Then, the partition function of the vertex model can be written as
\begin{equation}
    \mathcal{Z} = \sum_{\vec{s} \in \{-1,1\}^N} \prod_{p\in G} \sum_{b=0}^{V_p} \delta\left(S_{\text{tot}}(p) = 2b - V_p\right) W_p(b),
\end{equation}
where $S_{\text{tot}}(p) = \sum_{(p,q) \in G} s_{pq}$.  We perform a Fourier transform of the $\delta$-functions using the relation
\begin{equation}
    \delta \left(\sum_{i \in p} s_i = c\right) = \frac{1}{2\pi}\int_0^{2\pi} d\theta_p\, e^{i \theta_p \left(\sum_{i \in p} s_i - c\right)},
\end{equation}
and we may rearrange the partition function upon inserting this identity to obtain
\begin{align}
    \mathcal{Z} = \frac{1}{(2\pi)^n}\int_0^{2\pi} d^n\theta\sum_{\vec{s} \in \{-1,1\}^N} \prod_{(q,\ell)\in G_D} e^{i s_{q\ell} (\theta_q + \theta_\ell)} \prod_{p\in G_D} \sum_{b=0}^{V_p} e^{-i \theta_p (2b - V_p)} W_p(b).
\end{align}
Performing the sum over spins, we find
\begin{align}
    \mathcal{Z} = \frac{2^N}{(2\pi)^n}\int_0^{2\pi} d^n\theta \prod_{(q,\ell)\in G_D} \cos (\theta_q + \theta_\ell) \prod_{p\in G_D} \sum_{b=0}^{V_p} e^{-i \theta_p (2b - V_p)} W_p(b).
\end{align}
Next, expand the cosine using $\cos (\theta_q + \theta_\ell) = \cos \theta_q \cos \theta_{\ell} - \sin \theta_q \sin \theta_{\ell}$, and associate no bond with a factor of $\cos \theta_q \cos \theta_{\ell}$ and a bond with a factor of $- \sin \theta_q \sin \theta_{\ell}$.  Performing this expansion and regrouping terms maps the partition function to
\begin{equation}
    \mathcal{Z} = \sum_{k_1, k_2,\cdots, k_n} \frac{2^N}{(2\pi)^n} \int_0^{2\pi} d^n\theta \prod_{p \in G_D} \sum_{b=0}^{V_p} i^{V_p-k_p} \cos^{k_p} \theta_p \sin^{V_p - k_p} \theta_p  e^{-i \theta_p (2b - V_p)} W_p(b),
\end{equation}
which can be written in the compact form
\begin{equation}
    \mathcal{Z} = \sum_{k_1, k_2,\cdots, k_n} \prod_p W'_p(k_p),
\end{equation}
where
\begin{equation}\label{eqn:selfdualrel}
    W'_p(k_p) = \sum_{b=0}^{V_p} i^{V_p-k_p} \left \langle \cos^{k_p} \theta_p \sin^{V_p- k_p} \theta_p  e^{-i \theta_p (2b - V_p)} \right \rangle W_p(b).
\end{equation}
Here, the notation $\langle f(\theta) \rangle = \frac{1}{2\pi} \int_0^{2\pi} f(\theta)\,d\theta$.  This defines a new vertex model precisely with weights $W'$.  The new weights are related to the old weights by the linear map $\mathcal{M}\colon W_p \to W'_p$, which is defined on the space of vertex weights.  The matrix $\mathcal{M}$ has elements
\begin{equation}
    \mathcal{M}_{ab} = i^{V_p-a} \left \langle \cos^{a} \theta_p \sin^{V_p- a} \theta_p  e^{-i \theta_p (2b - V_p)} \right \rangle.
\end{equation}
$\mathcal{M}$ can be divided into disjoint eigenspaces corresponding to each of its distinct eigenvalues.  Any configuration of vertex weights which lives entirely in a given eigenspace will remain in the eigenspace under the application of $\mathcal{M}$.  Therefore, these eigenspaces define a self-dual manifold; if a parameterization of a vertex model pierces the self-dual manifold \emph{and} exhibits a single phase transition, then the transition point occurs at the intersection with the self-dual manifold.  An example of $\mathcal{M}_{ab}$ for $V_p=6$ is
\begin{equation}
\renewcommand{\arraystretch}{1.3}
    \mathcal{M}_{ab} = \begin{pmatrix}
\frac{1}{8} & -\frac{3}{4} & \frac{15}{8} & -\frac{5}{2} & \frac{15}{8} & -\frac{3}{4} & \frac{1}{8}\\
 -\frac{1}{8} & \frac{1}{2} & -\frac{5}{8} & 0 & \frac{5}{8} & -\frac{1}{2} & 
 \frac{1}{8}\\
 \frac{1}{8} & -\frac{1}{4} & -\frac{1}{8} & \frac{1}{2} & -\frac{1}{8} & -\frac{1}{4} & \frac{1}{8}\\
 -\frac{1}{8} & 0 & \frac{3}{8} & 0 & -\frac{3}{8} & 0 & \frac{1}{8}\\
 \frac{1}{8} & \frac{1}{4} & -\frac{1}{8} & -\frac{1}{2} & -\frac{1}{8} & \frac{1}{4} & \frac{1}{8}\\
 -\frac{1}{8} & -\frac{1}{2} & -\frac{5}{8} & 0 & \frac{5}{8} & \frac{1}{2} & 
 \frac{1}{8}\\
 \frac{1}{8} & \frac{3}{4} & \frac{15}{8} & \frac{5}{2} & \frac{15}{8} & \frac{3}{4} & \frac{1}{8}\\
\end{pmatrix},
\end{equation}
and explicitly, it can be seen that the eigenvalues of $\mathcal{M}_{ab}$ are $\pm 1$. The eigenspaces corresponding to these eigenvalues are
\begin{equation}
    \mathcal{V}_{-1} = \text{span}\left\{ \begin{pmatrix} -4\\0\\1\\-1\\0\\0\\1
    \end{pmatrix}, \begin{pmatrix} -12\\1\\4\\-3\\0\\1\\0
    \end{pmatrix}, \begin{pmatrix} -15\\0\\4\\-3\\1\\0\\0
    \end{pmatrix}\right\}
\end{equation}
and 
\begin{equation}
    \mathcal{V}_{1} = \text{span}\left\{ \begin{pmatrix} 4\\-2\\1\\0\\0\\0\\1
    \end{pmatrix}, \begin{pmatrix} 0\\-1\\0\\0\\0\\1\\0
    \end{pmatrix}, \begin{pmatrix} -15\\10\\-4\\0\\1\\0\\0
    \end{pmatrix}, \begin{pmatrix} -20\\10\\-4\\1\\0\\0\\0
    \end{pmatrix}\right\}.
\end{equation}
The vector $W_p(k) = \langle \cos^k \theta \sin^{V_p-k} \theta \rangle$, or $\langle 5,0,1,0,1,0,5 \rangle$ explicitly, lies in the $\mathcal{V}_1$ eigenspace, which we have argued in the text corresponds to a self-dual point.  However, what we point out here is that this is a self-dual point on \emph{any} 32-vertex model defined on a 6-coordinated lattice in \emph{any} dimension.  In a previous paper \cite{Balasubramanian}, we showed in general that the vertex weights 
\begin{equation}\label{eq:vertweightsgen}
    W_p(k) = \langle \cos^k \theta \sin^{V_p-k} \theta \rangle
\end{equation}
are a self-dual point for a vertex model defined on a $V_p$-coordinated lattice in any dimension.

The decorated Wegner duality can be seen by composing the Wegner duality with a more innocent duality.  Suppose we modify the partition function to have a dimer fugacity of $-1$; i.e. by adding an extra contribution to the Boltzmann weight equalling $(-1)^{\sum_i d_i}$ with $d_i = 1$ for a dimer on edge $i$ and $d_i = 0$ otherwise.  Consider the set of dimer configurations where two configurations are connected by sequence of loop moves (these states define a topological sector).  Each loop move conserves the parity of the number of dimers, and therefore the partition function within the topological sector is the same regardless of the extra factor $(-1)^{\sum_i d_i}$.  To see what this implies about the vertex weights, notice that since each dimer touches two vertices, a dimer contributes a factor of $i$ to each vertex weight.  Therefore, the vertex weights $\widetilde{W}_p(k) = (i)^{k} W_p(k)$.  For the diamond, cubic, and BCC lattices, the dual weights are
\begin{align}
    \widetilde{W}_4(0) &= \widetilde{W}_4(4) = u, \hspace{0.5cm}  \widetilde{W}_4(2) = -1 \\
    \widetilde{W}_6(0) &= -\widetilde{W}_6(6) = u, \hspace{0.5cm}  \widetilde{W}_6(4) = -\widetilde{W}_6(2) = 1 \\
    \widetilde{W}_8(0) &= \widetilde{W}_8(8) = u, \hspace{0.5cm}  \widetilde{W}_8(2) = \widetilde{W}_8(6) = -v \hspace{0.5cm}  \widetilde{W}_8(4) = 1.
\end{align}
Applying Wegner's duality to these modified weights yields the decorated Wegner duality.  Note that the decorated Wegner duality is only a self duality for the diamond and BCC lattices.  For the cubic lattice, it yields the fact that the gapless point ($u=5$) maps onto an ice rule-constrained model with 3 dimers at a site. 

\section{Further details of height field theory for vertex models}
With the representation from Eqn.~\ref{eq:vertweightsgen} in mind, it is possible to verify the claim that the partition function at the self-dual point can be written as
\begin{equation}
    \mathcal{Z} \propto \int_0^{2\pi} d^n \theta \prod_{\langle p, q \rangle} \cos\left(\theta_p - \theta_q\right) = \int_0^{2\pi} d^n \theta \prod_{\langle p, q \rangle} \left(\cos\theta_p \cos\theta_q + \sin\theta_p \sin\theta_q\right)
\end{equation}
by expanding the cosine term and regrouping.  In particular, we identify $\cos \theta_p \cos \theta_q$ as a dimer at edge $(p,q)$ and $\sin \theta_p \sin \theta_q$ as an empty link at at edge $(p,q)$, yielding a vertex model with weights
\begin{equation}
    W_p(k) = \langle \cos^k \theta \sin^{V_p-k} \theta \rangle.
\end{equation}
With nonzero $\alpha$, the vertex weights are
\begin{equation}
    W_p(k) = \langle (\cos \theta + \alpha \cos 3\theta)^k (\sin \theta - \alpha \sin 3\theta)^{V_p-k} \rangle
\end{equation}
which gives a relation between $u$ and $\alpha$ for the diamond lattice $V_p = 4$:
\begin{equation}
    u(\alpha) = 3 + \frac{16 \alpha}{\alpha^4 + 4 \alpha^2 - 4 \alpha + 1}
\end{equation}
For the cubic lattice with $V_p = 6$, we find
\begin{equation}
    u(\alpha) = 5 + \frac{40 \alpha (1+2\alpha^2)}{\alpha^6 + 9 \alpha^4 - 10 \alpha^3 + 9 \alpha^2 - 5 \alpha + 1}.
\end{equation}
For lattices with $V_p > 6$, other perturbations can be achieved via the deformation
\begin{equation}
    W_p(k) = \langle (\cos \theta + \alpha_n \cos n\theta)^k (\sin \theta - \alpha_n \sin n\theta)^{V_p-k} \rangle
\end{equation}
for $n \in 4 \mathbb{Z} - 1$.  For example, on the BCC lattice we have nontrivial weights for $n = 3, 7$, where $\alpha_{3} = \alpha$ and $\alpha_7 = \beta$. The functions $u(\alpha, \beta)$ and $v(\alpha, \beta)$, are ratios of complicated polynomials in $\alpha$ and $\beta$ and for brevity, we will not write down these functions.

There is an alternative way to understand what perturbing around the ice rule point does near the dual gapless point.  We start with the previously derived form of the partition function:
\begin{align}
    \mathcal{Z} = \frac{2^N}{(2\pi)^n}\int_0^{2\pi} d^n\theta \prod_{(q,\ell)\in G_D} \cos (\theta_q + \theta_\ell) \prod_{p\in G_D} \sum_{b=0}^{V_p} e^{-i \theta_p (2b - V_p)} W_p(b).
\end{align}
At the ice rule point, $2b - V_p = 0$, and perturbing about the ice rule point by selecting $W_{p}(b) = \alpha_b \ll 1$ for $b$ even and $0$ otherwise, we find
\begin{align}
    \mathcal{Z} = \frac{2^N}{(2\pi)^n}\int_0^{2\pi} d^n\theta \prod_{(q,\ell)\in G_D} \cos (\theta_q + \theta_\ell) \prod_{p\in G_D} \left(1 + \sum_{b=0, b \in 2\mathbb{Z}}^{V_p/2 - 1} 2\alpha_b \cos((2 b-V_p)\theta_p)\right).
\end{align}
And performing the sublattice symmetry $\theta_A \to -\theta_A$, we obtain
\begin{align}
    \mathcal{Z} = \frac{2^N}{(2\pi)^n}\int_0^{2\pi} d^n\theta \prod_{(q,\ell)\in G_D} \cos (\theta_q - \theta_\ell) \prod_{p\in G_D} \left(1 + \sum_{b=0, b \in 2\mathbb{Z}}^{V_p/2 - 1} 2\alpha_b \cos((2 b-V_p)\theta_p)\right).
\end{align}
As before, we may perform a gradient expansion writing $\cos(\theta_q - \theta_\ell) \approx 1 - \frac{K}{2} (\nabla \theta)^2$.  Furthermore, the cosine terms can also be treated perturbatively, giving
\begin{align}
    \mathcal{Z} \approx \frac{2^N}{(2\pi)^n}\int_0^{2\pi} d^n\theta \prod_{(q,\ell)\in G_D} \exp\left(-\frac{K}{2} \int d^3 x\,\left((\nabla \theta)^2 + \frac{4\alpha_0}{K} \cos(V_p \theta_p) + \frac{4\alpha_2}{K} \cos((V_p-4) \theta_p) + \cdots \right)\right),
\end{align}
which gives the right effective field theory for $V_p = 4, 8$.  Notably, for $V_p = 4$, adding a small weight for $W(0)$ and $W(4)$ perturbs into the ferromagnetic side of the gapless point.  

\section{Operators at and away from the gapless point}

In this section, we discuss the various field theoretic operators in the vertex model that were discussed in the main text, and how one might compute them in the microscopic vertex model.  We first start with the effective partition function:
\begin{equation}
\mathcal{Z} \propto \int_0^{2\pi} d^n \theta \prod_{\langle p, q \rangle} \cos\left(\theta_p - \theta_q\right)
\end{equation}
where, as previously discussed, $\cos \theta_p \cos \theta_q$ corresponds to a dimer on edge $(p,q)$ and $\sin \theta_p \sin \theta_q$ corresponds to an empty link.  Consider the quantity
\begin{equation}
    \left\langle \frac{\cos\left(\theta_{\bm{x}} + \theta_{\bm{x} + \bm{e_i}}\right)}{\cos\left(\theta_{\bm{x}} - \theta_{\bm{x} + \bm{e_i}}\right)} \frac{\cos\left(\theta_{\bm{y}} + \theta_{\bm{y} + \bm{e_j}}\right)}{\cos\left(\theta_{\bm{y}} - \theta_{\bm{y} + \bm{e_j}}\right)}\right\rangle = \frac{\int_0^{2\pi}  d^n \theta \, \cos\left(\theta_{\bm{x}} + \theta_{\bm{x} + \bm{e_i}}\right) \cos\left(\theta_{\bm{y}} + \theta_{\bm{y} + \bm{e_j}}\right) \prod_{\langle p, q \rangle} \cos\left(\theta_p - \theta_q\right)}{\int_0^{2\pi} d^n \theta \prod_{\langle p, q \rangle} \cos\left(\theta_p - \theta_q\right)}
\end{equation}
where the notation $\prod_{\langle p, q \rangle}$ takes a product over all edges, omitting the two edges at locations $\bm{x}$ and $\bm{y}$ with orientations $\bm{e_i}$ and $\bm{e_j}$.  Note that $\cos\left(\theta_{\bm{x}} + \theta_{\bm{x} + \bm{e_i}}\right)$ provides a weight of $-1$ to an empty link relative to a dimer.  Thus, this is simply a correlation function between dimer operators $\langle d_{\bm{x}} d_{\bm{y}} \rangle$.  Next, we consider the deconfinement order parameter presented in the main text:
\begin{equation}
    \left\langle \frac{\sin\left(\theta_{\bm{x}} + \theta_{\bm{x} + \bm{e_i}}\right)}{\cos\left(\theta_{\bm{x}} - \theta_{\bm{x} + \bm{e_i}}\right)} \frac{\sin\left(\theta_{\bm{y}} + \theta_{\bm{y} + \bm{e_j}}\right)}{\cos\left(\theta_{\bm{y}} - \theta_{\bm{y} + \bm{e_j}}\right)}\right\rangle = \frac{\int_0^{2\pi}  d^n \theta \, \sin\left(\theta_{\bm{x}} + \theta_{\bm{x} + \bm{e_i}}\right) \sin\left(\theta_{\bm{y}} + \theta_{\bm{y} + \bm{e_j}}\right) \prod_{\langle p, q \rangle} \cos\left(\theta_p - \theta_q\right)}{\int_0^{2\pi} d^n \theta \prod_{\langle p, q \rangle} \cos\left(\theta_p - \theta_q\right)}
\end{equation}
The numerator can be mapped onto a vertex model with two charge defects.  First expand $\sin\left(\theta_{\bm{x}} + \theta_{\bm{x} + \bm{e_i}}\right) = \sin \theta_{\bm{x}} \cos \theta_{\bm{x} + \bm{e_i}} + \cos \theta_{\bm{x}} \sin \theta_{\bm{x} + \bm{e_i}}$.  We arbitrarily choose a dimer to correspond to $\sin \theta_{\bm{x}} \cos \theta_{\bm{x} + \bm{e_i}}$ and an empty link to correspond to $\cos \theta_{\bm{x}} \sin \theta_{\bm{x} + \bm{e_i}}$.  This means that the vertex weights at $\bm{x}$ and $\bm{y}$ need to be modified.  Call $\ell$ the link where this defect is located.  Call $v$ a vertex configuration at $\bm{x}$ and $\overline{v}$ the vertex configuration where the value of the dimer at $\ell$ has been flipped.  The number of dimers in configuration $v$ will be denoted by $|v|$.  The new vertex weights, denoted by $W'_{V_p}$ are therefore
\begin{equation}
    W'_{V_p}(|v|) = W_{V_p}(|\overline{v}|)
\end{equation}
where $W_{V_p}(n) = \langle \cos^{n} \theta \sin^{V_p - n} \theta \rangle$.  Notice that the vertex weights $W'_{V_p}$ at sites $\bm{x}$ and $\bm{y}$ only allow an odd number of dimers.  As this explicitly breaks the even dimer constraint, one can view the numerator as a partition function with two defects inserted at $\bm{x}$ and $\bm{y}$.  Calling $\mathcal{Z}(\bm{x},\bm{y})$ the partition function with two defects located at $\bm{x}$ and $\bm{y}$, the desired correlation function is $\mathcal{Z}(\bm{x},\bm{y})/\mathcal{Z}$.

We may write the correlation function as the expectation value of a certain defect operator.  To do so, note that each term in the partition function in the numerator can be turned into an even dimer constrained configuration in the denominator by performing dimer flips along a path from $\bm{x}$ to $\bm{y}$.  Label the path from $\bm{x}$ to $\bm{y}$ with $\mathcal{P}$ where $\mathcal{P}(i)$ gives the $i$th site along the path.  For some defected configuration of dimers $C$, flipping dimers along $\mathcal{P}$ constructs an even dimer configuration in the denominator whose Boltzmann weight is $W_2(\overline{C})$ (the bar denotes flipping of dimers along the defect line in $C$).  The relation between the original weight $W_1(C)$ and $W_2(\overline{C})$ is
\begin{equation}
    \frac{W_1(C)}{W_2(\overline{C})} = \prod_{i \neq \bm{x}, \bm{y}} \eta_i,
\end{equation}
where $\eta_i = W_i(C_i)/W_i(\overline{C}_i)$, $C_i$ being the vertex configuration at $i$ and $W_i$ denoting the vertex weight at site $i$.  More specifically, this quantity is $1$ when $d_{i-1, i} d_{i, i+1} = -1$, i.e. one of the two edges along the path that touch $\mathcal{P}(i)$ have a dimer.  When $d_{i-1, i} d_{i, i+1} = 1$, $\eta_i$ is a ratio of $ W_i(|C_i|)/W_i(|\overline{C}_i|)$, where $|C_i|$ denotes the number of dimers touching $i$.  Let $D_i$ be the total number of dimers at $i$ that \emph{do not} belong to the path $\mathcal{P}$.  Therefore, we may write
\begin{equation}
    \left\langle \frac{\sin\left(\theta_{\bm{x}} + \theta_{\bm{x} + \bm{e_i}}\right)}{\cos\left(\theta_{\bm{x}} - \theta_{\bm{x} + \bm{e_i}}\right)} \frac{\sin\left(\theta_{\bm{y}} + \theta_{\bm{y} + \bm{e_j}}\right)}{\cos\left(\theta_{\bm{y}} - \theta_{\bm{y} + \bm{e_j}}\right)}\right\rangle = \left\langle \prod_{i \neq \bm{x}, \bm{y}} \eta_i \right\rangle = \left\langle \exp\left(\sum_i \log \frac{W(D_i -d_{i-1,i}-d_{i,i+1})}{W(D_i +d_{i-1,i}+d_{i,i+1})}\right) \right\rangle
\end{equation}
where in second equality, the expression in the summand can be written as a polynomial in the $d_{i, i+1}$ variables.  This quantity can be computed via Monte Carlo simulations to check the predicted theoretical behavior.

Finally, there is a dressed Wilson line-like operator which we expect to decay algebraically and not exhibit long range order.  This operator can be constructed by applying duality to the spin-spin correlators at the ice rule constrained point.  We compute:
\begin{align}
     \langle d_{p_1 q_1} d_{p_2 q_2}\rangle = \frac{1}{\mathcal{Z}} \sum_{\vec{s} \in \{-1,1\}^N} s_{p_1 q_1} s_{p_2 q_2} \prod_{p\in G} \sum_{b=0}^{V_p} \delta\left(S_{\text{tot}}(p) = 2b - V_p\right) W_p(b).
\end{align}
Expanding the $\delta$-functions as before gives the expression
\begin{align}
    \langle d_{p_1 q_1} d_{p_2 q_2}\rangle \propto \frac{1}{\mathcal{Z}}\int_0^{2\pi} d^n\theta \, \sin (\theta_{p_1} + \theta_{q_1}) \sin (\theta_{p_2} + \theta_{q_2}) \prod_{(q,\ell)\in G_D} \cos (\theta_q + \theta_\ell) \prod_{p\in G_D} \sum_{b=0}^{V_p} e^{-i \theta_p (2b - V_p)} W_p(b).
\end{align}
Setting $2b - V_p = 0$ to impose the ice rule constraint, we find
\begin{align}
    \langle d_{p_1 q_1} d_{p_2 q_2}\rangle \propto \frac{1}{\mathcal{Z}}\int_0^{2\pi} d^n\theta \, \sin (\theta_{p_1} + \theta_{q_1}) \sin (\theta_{p_2} + \theta_{q_2}) \prod_{(q,\ell)\in G_D} \cos (\theta_q + \theta_\ell),
\end{align}
and a sublattice transformation $\theta_p \to -\theta_p$ on one of the sublattices gives the identity
\begin{align}
    \langle d_{\bm{x}, \bm{x} + \bm{e}_i} d_{\bm{y}, \bm{y} + \bm{e}_j}\rangle = \left\langle \frac{\sin\left(\theta_{\bm{x}} - \theta_{\bm{x} + \bm{e_i}}\right)}{\cos\left(\theta_{\bm{x}} - \theta_{\bm{x} + \bm{e_i}}\right)} \frac{\sin\left(\theta_{\bm{y}} - \theta_{\bm{y} + \bm{e_j}}\right)}{\cos\left(\theta_{\bm{y}} - \theta_{\bm{y} + \bm{e_j}}\right)}\right\rangle.
\end{align}
As before, we expand $\sin\left(\theta_{\bm{x}} + \theta_{\bm{x} + \bm{e_i}}\right) = \sin \theta_{\bm{x}} \cos \theta_{\bm{x} + \bm{e_i}} - \cos \theta_{\bm{x}} \sin \theta_{\bm{x} + \bm{e_i}}$, which adds a dimer fugacity of $-1$ for a dimer relative to no dimer.  Therefore, this object is just the dressed Wilson line corresponding to $\langle \sin (2 \theta(\bm{x})) \sin (2 \theta(\bm{y})) \rangle$ decorated on its ends with a dimer operator charged under the global $\mathbb{Z}_2$ symmetry:
\begin{equation}
    \left\langle \frac{\sin\left(\theta_{\bm{x}} - \theta_{\bm{x} + \bm{e_i}}\right)}{\cos\left(\theta_{\bm{x}} - \theta_{\bm{x} + \bm{e_i}}\right)} \frac{\sin\left(\theta_{\bm{y}} - \theta_{\bm{y} + \bm{e_j}}\right)}{\cos\left(\theta_{\bm{y}} - \theta_{\bm{y} + \bm{e_j}}\right)}\right\rangle = \left\langle (-1)^{d_{\bm{x}, \bm{x} + \bm{e_i}}}\exp\left(\sum_i \log \frac{W(D_i -d_{i-1,i}-d_{i,i+1})}{W(D_i +d_{i-1,i}+d_{i,i+1})} \right)(-1)^{d_{\bm{y}, \bm{y} + \bm{e_j}}} \right\rangle
\end{equation}
Our expectation is that this quantity has power law decay in $|\bm{x} - \bm{y}|$.    Notice that in addition to these operators, there are other kinds of operators of the form $\cos m \theta$ and $\sin n \theta$; the former corresponds to a local operator, while the latter corresponds to a  certain kind of Wilson line.  The effective field theory provided in the main text gives simple and sharp predictions for expectation values of these operators; in particular, they all exhibit long range order at the gapless point and may result in unusual coexistence phases away from the gapless point. 

Next, we discuss how these operators get perturbed away from the gapless point.  We utilize the following identity, presented in the main text:
\begin{equation}
    \mathcal{Z}_{\alpha} = \int_0^{2\pi} d^n\theta \prod_{\langle p,q \rangle} \left(f(\theta_p)f(\theta_q) + g(\theta_p)g(\theta_q)\right),
\end{equation}
where $f(\theta) = \cos \theta + \alpha \cos(3\theta)$ and $g(\theta) = \sin \theta - \alpha \sin(3\theta)$, where we have turned off the $\beta$ term.  Since $f(\theta_p)f(\theta_q)$ is now associated with a dimer and $g(\theta_p)g(\theta_q)$ an empty link, the modified dimer order parameter is
\begin{equation}
d_{\bm{x}}(\alpha) = \frac{f(\theta_{\bm{x}}) f(\theta_{\bm{x}+\bm{e_i}}) - g(\theta_{\bm{x}}) g(\theta_{\bm{x}+\bm{e_i}})}{f(\theta_{\bm{x}}) f(\theta_{\bm{x}+\bm{e_i}}) + g(\theta_{\bm{x}}) g(\theta_{\bm{x}+\bm{e_i}})}.
\end{equation}
We may define a deconfinement parameter in an analogous way as was done at the gapless point:
\begin{equation}
\mathcal{O}_{\bm{x}}(\alpha) = \frac{f(\theta_{\bm{x}}) g(\theta_{\bm{x}+\bm{e_i}}) + g(\theta_{\bm{x}}) f(\theta_{\bm{x}+\bm{e_i}})}{f(\theta_{\bm{x}}) f(\theta_{\bm{x}+\bm{e_i}}) + g(\theta_{\bm{x}}) g(\theta_{\bm{x}+\bm{e_i}})}.
\end{equation}
Under Wegner's duality, $\mathcal{O}_{\bm{x}}(\alpha)$ maps to the dual value of the dimer density $d_{\bm{x}}(-\alpha)$.  These operators can be expressed on the lattice in an analogous way to the above derivation at the gapless point.  When nonzero $\beta$ is added, the dimer and deconfinement operators become $d(\alpha, \beta)$ and $\mathcal{O}(\alpha, \beta)$ and under duality, $\mathcal{O}(\alpha, \beta)$ is mapped to $d(-\alpha, \beta)$.  All of these results generalize in a straightforward manner when $V_p > 8$.

Similarly, one can construct an order parameter that generalizes the deconfinement parameter decorated by $\mathbb{Z}_2$ charges by defining
\begin{equation}
\mathcal{Q}_{\bm{x}}(\alpha) = \frac{f(\theta_{\bm{x}}) g(\theta_{\bm{x}+\bm{e_i}}) - g(\theta_{\bm{x}}) f(\theta_{\bm{x}+\bm{e_i}})}{f(\theta_{\bm{x}}) f(\theta_{\bm{x}+\bm{e_i}}) + g(\theta_{\bm{x}}) g(\theta_{\bm{x}+\bm{e_i}})}.
\end{equation}
using the explicit forms of $f$ and $g$, one can verify that even away from the gapless point, the leading order behavior is $\langle \mathcal{Q}_{\bm{x}}(\alpha) \mathcal{Q}_{\bm{y}}(\alpha) \rangle \propto \langle \nabla \theta(\bm{x}) \nabla \theta(\bm{y}) \rangle$ which decays exponentially in all gapped phases.  Thus, this order parameter only serves as a diagnostic for critical behavior, and does not distinguish confined and deconfined phases of the gauge theory.  This conclusion still holds at nonzero $\beta$.

Finally, a more traditional operator that one computes is a Wilson loop operator of the dimer variables $W_C \sim \left\langle \prod_{i \in C} d_i \right\rangle$.  At the gapless point, we expect a universal scaling law first predicted by Peskin: if the Wilson loop has width $w$ and length $L$, and $L$ is taken to be sufficiently larger than $w$, then $W_C \sim e^{-k L/w}$ for some constant $k$ \cite{peskin1980critical}.

\section{Field theory dualities for Wegner and decorated Wegner duality}

Here, we note that the lattice dualities that we have studied are related to standard field theoretic dualities.  From the field theory perspective, the mapping from an ice rule model onto the self dual point ($u = 0$ to $u = 3$ in the case of the diamond lattice) is a consequence of the well-known mapping from a pure gauge theory to a compact boson (called the ``dual photon'' in standard field theory literature).  In particular, in 3+0D, the ice rule constraint enforces a divergence-free constraint on the electric field lines.  In the path integral representation, this can be written as
\begin{equation}
    \mathcal{Z} = \int \mathcal{D} \bm{E} \, \exp\left(-\frac{K}{2} \int d^3x\,\bm{E}^2\right) \delta(\nabla \cdot \bm{E} = 0).
\end{equation}
For simplicity, we set $K = 1$ for the remainder of this appendix.  Writing the $\delta$-function in terms of a Lagrange multiplier, we find that
\begin{equation}
    \mathcal{Z} = \int \mathcal{D} \bm{E} \, \mathcal{D} \theta \, \exp\left(-\frac{1}{2} \int d^3x\,\bm{E}^2 - i \theta (\nabla \cdot \bm{E})\right),
\end{equation}
and integrating over $\bm{E}$ maps us precisely onto the proposed action for the height fields $\theta$.  A similar analysis can be done in 2+1D by defining $\bm{E}$ in terms of electric and magnetic fields so that the divergence-free constraint is equivalent to Faraday's law.

Notice that the $\theta$ field is responsible for creating defects in the ice rule limit.  In particular, $\mathcal{M}(\bm{x}_0) = e^{-i \theta(\bm{x}_0)}$ creates a monopole at $\bm{x}_0$, as this corresponds to changing the Gauss law constraint to $\nabla \cdot \bm{E} = \delta^3(\bm{x} - \bm{x}_0)$ (on the lattice, this corresponds to enforcing an odd dimer constraint at $\bm{x}_0$).  Similarly, $\mathcal{M}^{\dagger}(\bm{x}_0) = e^{i \theta(\bm{x}_0)}$ creates an anti-monopole at $\bm{x}_0$.  As derived in the main text, the relevant perturbation driving the system away from the gapless point is $\alpha \cos(4\theta)$, will results in the following effective field theory when perturbed away from the spin ice point:
\begin{equation}
    \mathcal{Z} = \int \mathcal{D} \bm{E} \, \mathcal{D} \theta \, \exp\left(-\frac{1}{2} \int d^3x\,\bm{E}^2 - i \theta (\nabla \cdot \bm{E}) - \alpha \cos(4\theta)\right).
\end{equation}
Therefore, the transition out of the spin ice point corresponds to condensing a charge conjugation-symmetric monopole operator $\frac{1}{2} (\mathcal{M}(\bm{x})\mathcal{M}(\bm{x}) + \mathcal{M}^{\dagger}(\bm{x})\mathcal{M}^{\dagger}(\bm{x}))$.  When $\theta$ is pinned to the vacuum expectation value, fluctuations give rise to a mass term:
\begin{equation}
    \mathcal{Z} = \int \mathcal{D} \bm{E} \, \mathcal{D} (\delta \theta) \, \exp\left(-\frac{1}{2} \int d^3x\,\bm{E}^2 - i \frac{\pi}{4} (\nabla \cdot \bm{E}) - i (\delta \theta) (\nabla \cdot \bm{E}) + 8 \alpha (\delta \theta)^2\right).
\end{equation}
The second term is a boundary term corresponding to the enclosed electric flux, while the integral over $\delta \theta$ can be done explicitly to give
\begin{equation}
    \mathcal{Z} = \int \mathcal{D} \bm{E} \, \exp\left(-\frac{1}{2} \int d^3x\,\left(\bm{E}^2 + \frac{(\nabla \cdot \bm{E})^2}{32 \alpha}\right)\right),
\end{equation}
which therefore gaps out the theory.

To understand the Wegner duality, we noted in the main text that implementing duality on the lattice is equivalent to mapping $\alpha \to -\alpha$ in the partition function.  This can be verified by direct calculation.  Therefore, the operator $\cos 4\theta$ is duality odd.  For the BCC lattice, one can show that the choice of weights
\begin{equation}
    W_8(k) = \langle (\cos \theta + \beta \cos 7\theta)^k (\sin \theta - \beta \sin 7\theta)^{V_p-k} \rangle
\end{equation}
resides in the self dual boundary and is therefore duality even.  In the effective field theory, this corresponds to adding the operator $\cos 8\theta$.  Similarly, one can show that the choice of weights
\begin{equation}
    W_{12}(k) = \langle (\cos \theta + \gamma \cos 11\theta)^k (\sin \theta - \gamma \sin 11\theta)^{V_p-k} \rangle
\end{equation}
maps as $\gamma \to -\gamma$ under duality and is thus duality odd.  The corresponding operator in the field theory is $\cos 12\theta$  This pattern can be verified analytically (with the aid of a computer algebra system) to large values of $V_p$ indicating that $\cos n \theta$ for $n \in 8\mathbb{Z}+4$ is duality odd and $\cos n \theta$ for $n \in 8\mathbb{Z}$ is duality even.  This is consistent with the fact that duality maps $\theta \to \theta + \pi/4$.

Finally, we may compose the decorated Wegner duality and the Wegner duality to get a third duality.  For example, on the pyrochlore lattice, the decorated Wegner duality maps the interval $[0,1]$ to $[1,3]$ and the Wegner duality maps the interval $[1,3]$ to $[3,\infty)$.  Composing the two dualities maps $[0,1]$ to $[3,\infty)$.

\section{Detailed analysis of BCC lattice phase diagram}

In this section, we provide an in-depth analysis of the BCC lattice vertex model, whose pertinent features were discussed in the main text.  The BCC lattice is dual to a lattice of corner-sharing cubes, and the spins are located on the sites of this lattice (corresponding to dimer variables on the edges of the BCC lattice).  The vertex model on the corresponding lattice has $V_p = 8$.  Here, the phase diagram is parameterized by two variables, $u = W_8(8)/W_8(4)$ and $v = W_8(6)/W_8(4)$ (we will ignore the subscript $8$ for the remainder of the section), and possesses the same $\mathbb{Z}_2$ symmetry as before.

This lattice admits both a Wegner and decorated Wegner duality.  Wegner's duality gives
\begin{equation}
    u^* = \frac{35+u+28v}{3+u-4v}, \hspace{0.5cm} v^* = \frac{-5+u+4v}{3+u-4v}.
\end{equation}
In particular, the gapless point  $W(n) = \langle \cos^{8-n} \theta \sin^n \theta \rangle$ has the weights
\begin{equation}
    W(0) = W(8) = \frac{35}{3}, \hspace{0.3cm} W(2) = W(6) = \frac{5}{3}, \hspace{0.3cm} W(4) = 1.
\end{equation}
as discussed in the main text.  Wegner's duality gives a \emph{line} of self-dual points which is a solution to $u - 4v = 5$.  A significant portion of this self-dual line can be achieved by selecting the one-parameter family of vertex weights
\begin{equation}
    W(n, \beta) = \langle (\cos \theta + \beta \cos 7\theta)^n (\sin \theta - \beta \sin 7\theta)^{8-n} \rangle.
\end{equation}
corresponding to the duality even operator $\cos 8\theta$ in the effective field theory.  We may also access weights off of this self dual line through
\begin{equation}
    W(n, \alpha) = \langle (\cos \theta + \alpha \cos 3\theta)^n (\sin \theta - \alpha \sin 3\theta)^{8-n} \rangle.
\end{equation}
corresponding to the duality odd operator $\cos 4\theta$ in the effective field theory.  Upon perturbing about the gapless point where $u = 35/3$ and $v = 5/3$, the effective action for the model can be written as
\begin{equation}
   \mathcal{Z} = \int \mathcal{D}\theta(x)\, \exp\left(-\frac{K}{2}\int d^d x\, (\nabla \theta)^2 - \frac{4\alpha}{K} \cos (4 \theta) - \frac{4\beta}{K} \cos (8 \theta)\right),
\end{equation}
as evinced in the main text.  The behavior of this field theory for $d = 2$ and $d = 3$ are markedly different.  In 2+0D, the term $\cos(n\theta)$ can be irrelevant or relevant depending on $K$ and $n$, so the phase diagram might depend on more microscopic details.   Secondly, there is no regular 2D lattice with 8-fold rotation symmetry; there can be ``mixed'' lattices where some of the sites are 8-valent and some are not, but for these lattices the field theory example may no longer be accurate.

\begin{figure*}
    \centering
    \includegraphics[scale=0.4]{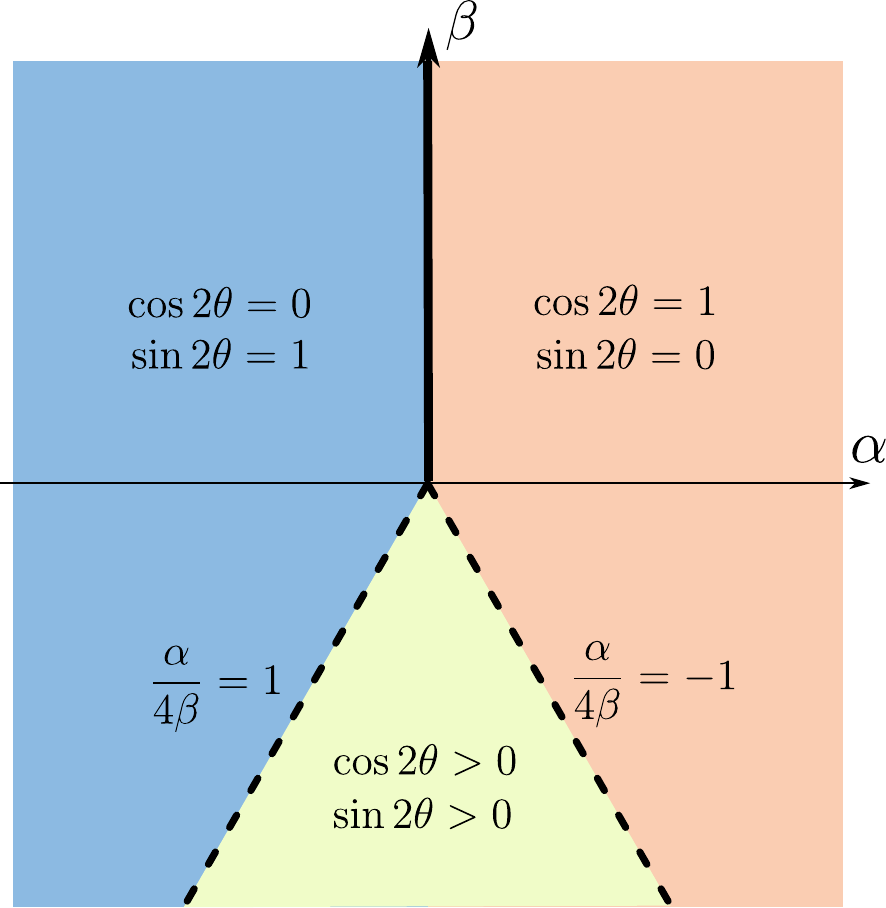}
    \caption{Mean field phase diagram for the BCC lattice vertex model.  Solid lines indicate first order transitions and dashed lines indicate second order transitions in the 3D Ising universality class.}
    \label{fig:BCC}
\end{figure*}

However, in 3+0D, all terms of the form $\cos(n \theta)$ are RG relevant and we expect that they might compete in an interesting way.  To simplify the analysis, we search for mean field solutions to the above action, corresponding to maximizing the potential
\begin{equation}
    V(\theta) = 4\alpha \cos (4 \theta) + 4\beta \cos (8 \theta) = 8\beta\left(\cos (4 \theta) + \frac{\alpha}{4 \beta}\right)^2 - c(\alpha, \beta)
\end{equation}
where $c(\alpha, \beta)$ is an overall constant that we ignore.  To optimize this potential, we discuss the following cases:
\begin{itemize}
    \item $\beta > 0$: when $\beta > 0$, we maximize $\left|\cos (4 \theta) + \frac{\alpha}{4 \beta}\right|$.  This splits into two cases: when $\alpha > 0$ then $\cos (4 \theta)$ is pinned to $1$ (i.e. the dimer order is $1$ and the deconfinement parameter is $0$).  When $\alpha < 0$, then $\cos (4 \theta)$ is pinned to $-1$ (i.e. the dimer order is $0$ and the deconfinement parameter is $1$).
    \item $\beta < 0$, $|\alpha/(4\beta)| > 1$: when $\beta < 0$, we minimize $\left|\cos (4 \theta) + \frac{\alpha}{4 \beta}\right|$.  For $|\alpha/(4\beta)| > 1$ and $\alpha > 0$, then $\cos (4 \theta)$ is pinned to $1$.  For $|\alpha/(4\beta)| > 1$ and $\alpha < 0$, then $\cos (4 \theta)$ is pinned to $-1$.
    \item $\beta < 0$, $|\alpha/(4\beta)| < 1$:  we still minimize $\left|\cos (4 \theta) + \frac{\alpha}{4 \beta}\right|$, but now the minimum value satisfies is $\cos(4 \theta) = -\alpha/(4\beta)$.
\end{itemize}
A diagram of the mean field solution is illustrated in Figure~\ref{fig:BCC}.  In particular, for $\beta < 0$, $|\alpha/(4\beta)| < 1$, this corresponds to a phase where both the dimer order parameter and the deconfinement parameter have nonzero vacuum expectation values (at least for sufficiently small values of $\alpha$ and $\beta$).  If we were to include fluctuations, this would give a mass term to lowest order, indicating that all three phases are gapped.

Next, we analyze the nature of the three critical lines in this model: $\beta > 0$ and $\alpha = 0$, $\beta < 0$ and $\alpha/(4\beta) = 1$, and $\beta < 0$ and $\alpha/(4\beta) = -1$.  First, note that the first transition line is self dual, while the second and third transition lines map onto each other under Wegner's duality.  The line $\beta > 0$ and $\alpha = 0$ is a first order transition.  The other two lines are more interesting: in the vicinity of these critical lines, expanding $\theta \approx \theta_0 + \delta\theta$ where $\theta_0$ is the mean field solution, we find
\begin{equation}
    S = \int d^3x\,\left(-\frac{K}{2}(\nabla \theta)^2 + 2\beta (\cos (4 \theta) \pm 1)^2\right) \approx \int d^3x\,\left(-\frac{K}{2}(\nabla (\delta\theta))^2 + 16\beta (\delta\theta)^4 \right)
\end{equation}
and deviations away from this critical line turns on a small mass term $m (\delta \theta)^2$.  This is simply the effective action near the Wilson-Fisher fixed point, which has critical exponents in the 3D Ising universality class.  We note that the slopes of the critical lines as predicted by mean field theory are not accurate, especially when compared to numerical simulations (see main text).

We now consider the decorated Wegner duality, whose action on the vertex weights is
\begin{equation}
    u^* = \frac{35+u-28v}{3+u+4v}, \hspace{0.5cm} v^* = -\frac{5-u+4v}{3+u+4v}.
\end{equation}
It is clear that the gapless point $(35/3, 5/3)$ maps onto the ice rule point $(0,0)$.  The region $\beta < 0$ maps onto $u < 0$ which is unphysical.  Therefore, the unusual critical line near the gapless point does not occur proximate to the spin ice point.  Under this duality, the first order transition line $\beta > 0$ and $\alpha = 0$ line maps onto $v = 0$ and $u < 1$.  This suggests that in the vicinity of the ice rule point, we expect there to be a $\mathbb{Z}_2$ deconfined phase and no other phases.

We make one more observation about the gapless multicritical point.  Note that this point is the result of the first order transition line meeting a second order transition line, and thus one might conclude that it is a tricritical point.  However, because the gapless point is dual to the RK point of a quantum spin ice model, we expect correlation functions to be equal to those of a free boson in 3D.  If the gapless point were tricritical, there would necessarily be additional logarithmic corrections in 3D.  Thus, there is apparently enough fine-tuning in this model to prevent the gapless point from being tricritical. 

\subsection{Spin model mappings for BCC lattice}

\begin{figure*}
    \centering
    \includegraphics[scale=0.4]{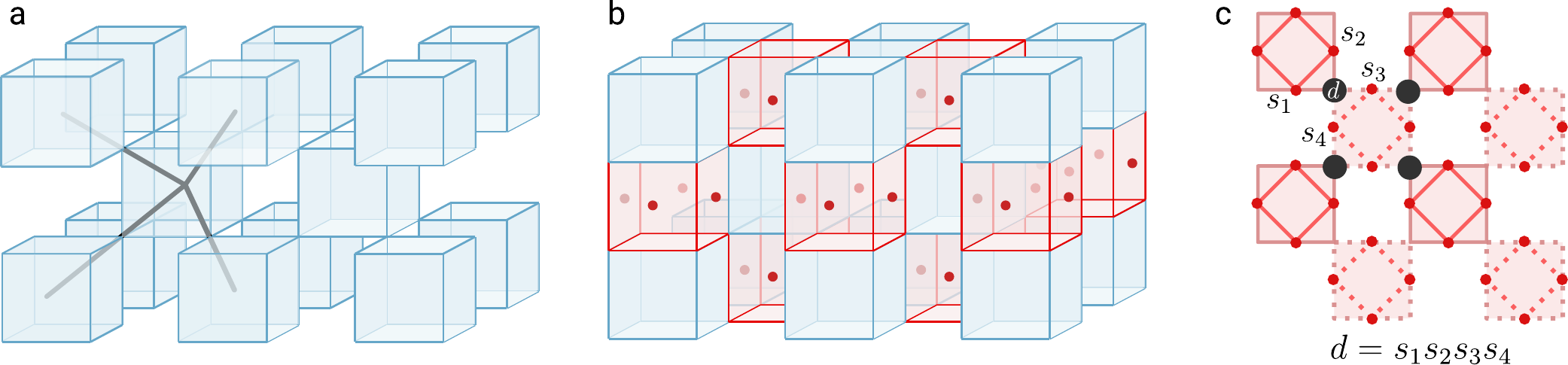}
    \caption{In panel (a), the dimers are shown, which connect cube centers.  In panel (b), the locations of the spins are shown which form a lattice corresponding to staggered layers of a square-octagon lattice.  In panel (c), a bird's eye view of two layers of the lattice are shown, and the assignment between a dimer (black dot) and four spins is indicated.}
    \label{fig:cubes}
\end{figure*}

\begin{figure*}
    \centering
    \includegraphics[scale=0.55]{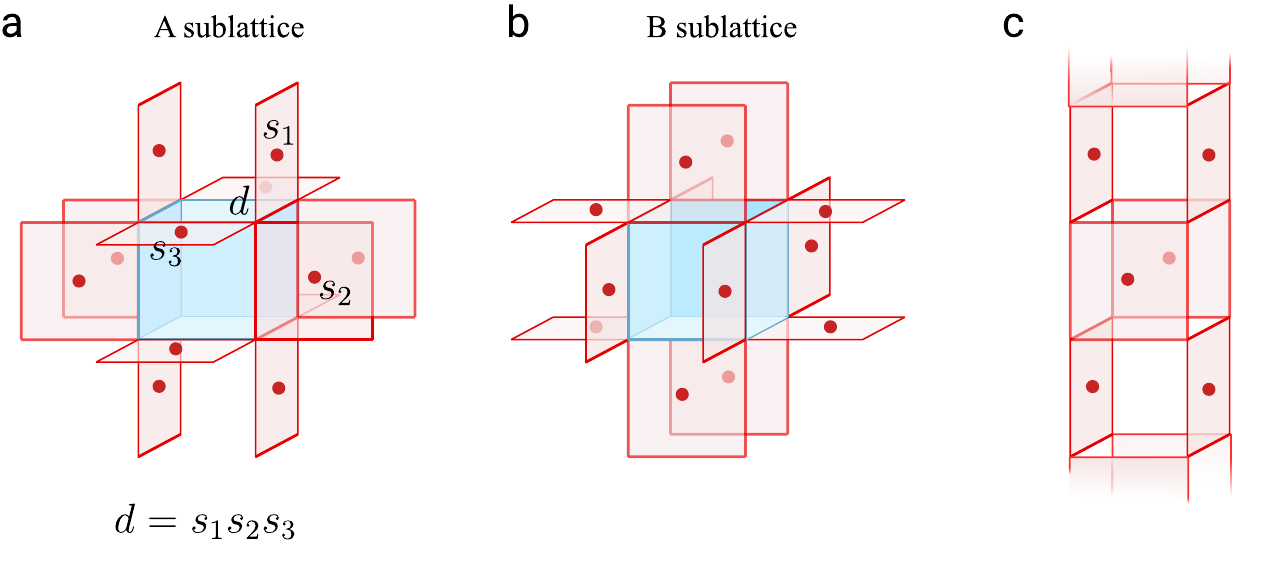}
    \caption{In panel (a), the locations of the spins in the A and B sublattice cubes are shown, and further the dimer-spin relation is also shown.  In panel (b), a ``tube''-like object is shown: the subsystem symmetry action corresponds to a product of $X$ along all spins lying on this tube.}
    \label{fig:cubes2}
\end{figure*}

In this subsection, we will discuss the construction of a classical spin model equivalent to the classical BCC lattice vertex model to better understand the interplay of global and gauge symmetries.  On the dual lattice of corner-sharing cubes, the dimers are located at the corners of the cubes, as shown in Fig.~\ref{fig:cubes}[a].  As in the main text, we want to represent the dimer variables in terms of effective spins which immediately satisfy the even dimer constraint by design.  One way to do so is to choose spins to be the red dots as shown in Fig.~\ref{fig:cubes}[b].  One can assign each dimer to the four spins it is nearest to.  More precisely, the lattice of the spins is equivalent to staggered layers of a square-octagon lattice: a bird's eye view is shown in Fig.~\ref{fig:cubes}[c]; in particular one layer of spins is shown using solid lines and the layer above is shown using dashed lines.  This staggering is repeated to form the lattice of spins.  In the bird's eye view, the dimers stick out of the page and are represented by black dots; the identification of a dimer with its four neighboring spins is shown.  Therefore, for each corner $c$ shared between two cubes, define the dimer variables $d_c = \prod_{i \in S} s_i^{(c)}$, where $S$ denotes the set of four nearby spins assigned to the dimer.  As before, $d_c = 1$ indicates an occupied dimer while $d_c = -1$ indicates an empty dimer.  We have the relation
\begin{equation}
    \prod_{c \in \mbox{\mancube}} d_c = \prod_{- \in \mbox{\mancube}} s_i^2 = 1,
\end{equation}
around each cube, which faithfully reproduces the even dimer constraint.  The partition function is $Z_{\text{spin}} = \sum_{\vec{s}} \exp\left(\sum_{\mbox{\mancube}} \mathcal{H}_{\mbox{\mancube}}\right)$, and the Hamiltonian is given by
\begin{equation}
    \mathcal{H}_{\mbox{\mancube}} = J_1 \left(\sum_{c \in \mbox{\mancube}} d_c\right)^2 + J_2 \left(\sum_{c \in \mbox{\mancube}} d_c\right)^4.
\end{equation}
We first note that there is a global $\mathbb{Z}_2$ symmetry corresponding to $d_c \to -d_c$, as well as a gauge symmetry corresponding to the classical Hamiltonian commuting with $\prod_{i \in \diamond} X_i$, where $\diamond$ indicates a square on the square octagon lattice.  Note that in this case, a ghost spin cannot be added because the vertex model is tuned by two parameters, not one.  For 4 dimers, $\sum_{c \in \mbox{\mancube}} d_c = 0$ and the Boltzmann weight $\exp\left(\mathcal{H}_{\mbox{\mancube}}\right) = 1$.  For 2 and 6 dimers, $\sum_{c \in \mbox{\mancube}} d_c = \pm 4$ and for 0 and 8 dimers, $\sum_{c \in \mbox{\mancube}} d_c = \pm 8$. Therefore, we solve for the vertex weights in terms of $J_1$ and $J_2$:
\begin{align}\label{eqn:couplings}
    u &= W(0) = W(8) = \exp\left(64 J_1 + 4096 J_2\right), \nonumber \\ v &= W(2) = W(6) = \exp\left(16 J_1 + 256 J_2\right),
\end{align}
and $W(4) = 1$.  At the self dual point, we find
\begin{equation}
    e^{J_1} = \left(\frac{5}{3}\right)^{5/64} \left(\frac{1}{7}\right)^{1/192} \hspace{0.5cm} e^{J_2} = \left(\frac{189}{125}\right)^{1/3072}.
\end{equation}
Surprisingly, there are alternative schemes corresponding to different assignments of the dimer variables to spins.  These result in new spin models with a similar phase diagram, but where the symmetries appear to be different.  One example of an alternative scheme is to associate a dimer with \emph{three} spins: the lattice of spins is shown in the Figure~\ref{fig:cubes2}[a], along with the assignment of the dimer and spin variables.  
As before, the constraint $\prod_{c \in \mbox{\mancube}} d_c = 1$ is immediately satisfied, and the global symmetry is $\prod_i X_i$ where $i$ ranges over all of the classical spins.  However, there is no longer a gauge constraint; instead one can construct subsystem symmetry operators given by a product of $X_i$ along the `tubes' depicted in Figure~\ref{fig:cubes2}[b].  The phase diagram for this model is expected to coincide with that of the spin model with the gauge constraint with the precise identification of phases being relegated to future work.  Given that the gauge constraint spin model supports a phase where symmetry breaking and deconfinement coexist, it seems that the subsystem symmetry spin model supports a phase where global and subsystem symmetry breaking coexist and are mixed in a nontrivial way.

\section{Matching degrees of freedom in spin and vertex models}

We show explicitly the correctness of the dimer model to spin model mapping presented in the main text by arguing that each dimer configuration is uniquely mapped to a spin configuration.  This can be shown by matching dimer and spin degrees of freedom through a counting argument.

We first work with the cubic lattice spin model, where
\begin{equation}
   \mathcal{Z}_{\text{spin}} = \sum_{\vec{s} \in \{-1,1\}^N} \prod_{\mbox{\mancube}} W_{\mbox{\mancube}} (\vec{s}). 
\end{equation}
First note the standard  $\mathbb{Z}_2$ gauge symmetry $P = \prod_{i \in v} X_i = X_1X_2X_3X_4X_5X_6$ where $X_i$ is the spin flip operator, and $i \in v$ denotes spins on the edges connected to vertex $v$.  Therefore, the number of gauge equivalent spin configurations is $2^N$, where $N$ is the number of lattice sites in the lattice.  The number of valid 32-vertex configurations $D$ is $2^{2N}$ for any lattice with coordination number 6.  If each dimer configuration corresponds to $2^{N}$ spin configurations, then there must be a total of $D \cdot 2^{N} = 2^{3N}$ spin configurations, which coincides with the number of spin degrees of freedom in the dual spin model (as spins are defined on the edges of the cubic lattice).  Therefore, the dimer model and spin model partition functions are related by $\mathcal{Z}_{\text{spin}}(J) = 2^{N} \mathcal{Z}_{\text{dimer}}(J)$.

Next, we show that the partition function for the 7-spin model maps to the partition function of the vertex model for the diamond lattice model.  The number of valid vertex model configurations on the diamond lattice is $D = 2^{N_T}$, where $N_T$ is the number of tetrahedra in the dual pyrochlore lattice.  The number of objects $\mathcal{O}$ (see main text) is $N_T$ and each object $\mathcal{O}$ is associated with a local gauge transformation.  The total number of spin degrees of freedom is therefore $D \cdot 2^{N_T} = 2^{2N_T} = 2^{N_H}$ where $N_H$ is the number of hexagons; this coincides with the number of degrees of freedom in the above spin model.  Therefore $\mathcal{Z}_{\text{spin}}(J) = 2^{N_T} \mathcal{Z}_{\text{dimer}}(J)$.

\begin{figure}
    \centering
    \includegraphics[scale=0.25]{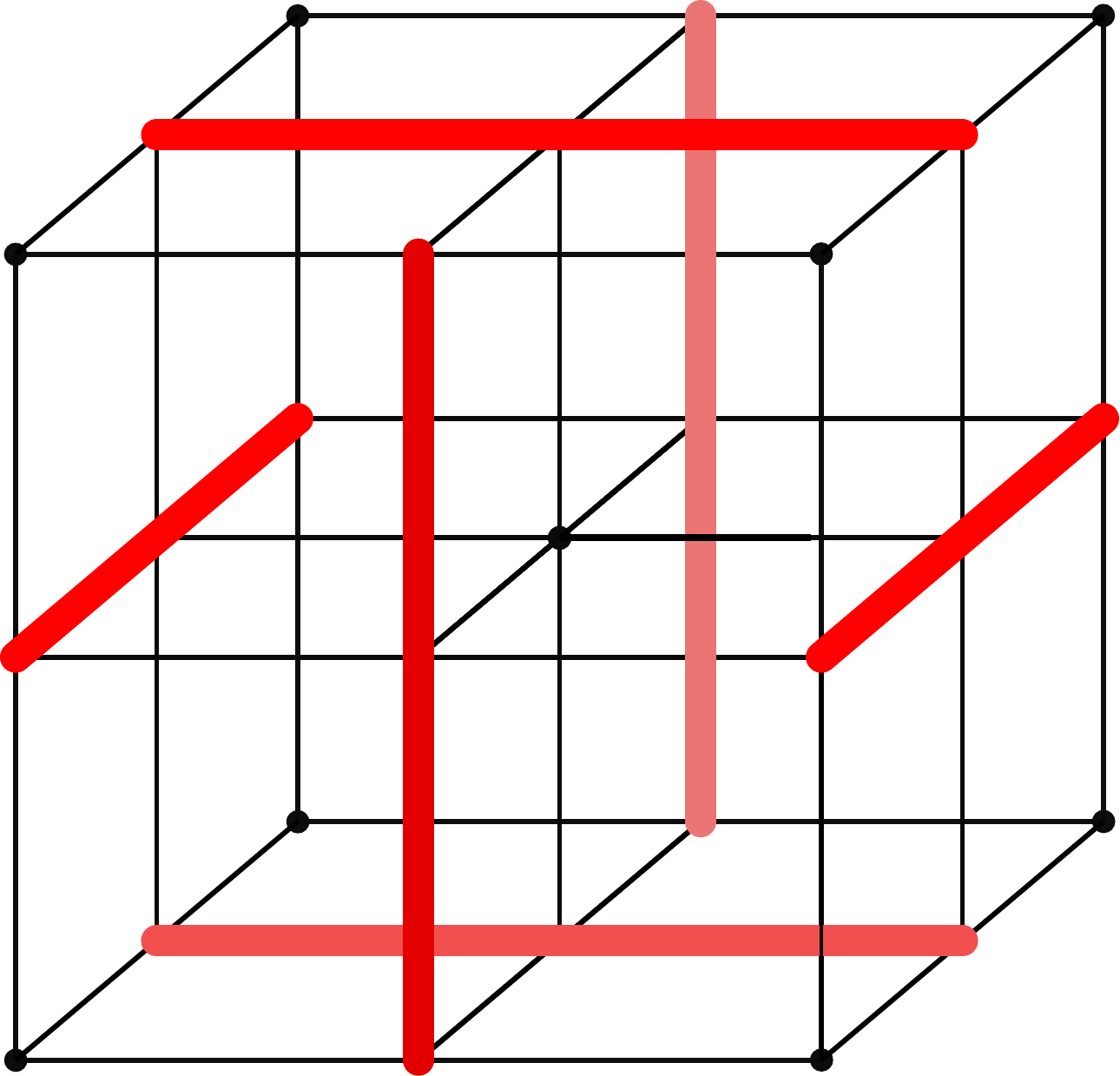}
    \caption{The configuration of spins that must be flipped to implement the global symmetry transformation.  The bond spins live on the links of the lattice.}
    \label{fig:bond}
\end{figure}

Furthermore, in the main text, we noted that the equivalent spin model possesses a global symmetry corresponding to $\tau \to -\tau$ and $d \to -d$.  It is not immediately obvious if there exists a transformation on the bond spins $s$ such that $d \to -d$.  We show this transformation in Figure~\ref{fig:bond}.  The red bonds indicate that the associated spins are flipped, and the structure of red bonds is repeated over the entire lattice.

\section{Even-odd effect for cubic lattice model}

(In this section we will change variables from $\mathcal{J}$ to $\beta$ for clarity).  We will discuss the phases of the gauge theory and the even-odd effect.  We start with the Hamiltonian for the 5-spin classical Hamiltonian describing the cubic lattice spin model.  Consider the Wilson loop operator
\begin{equation}
    W_{C} = \left \langle \prod_{i \in C} s_i \right \rangle,
\end{equation}
where $C$ is a closed loop containing the gauge spins.  Let us first analyze this quantity at low temperatures, in the deconfined phase.  Here, the standard analysis holds, which we review for convenience \cite{Kogut}.  We gauge fix so that the ground state spin configuration is the ferromagnetic configuration.  We then proceed with a low temperature expansion: flipping a gauge spin will negate 8 plaquette terms, while flipping a ghost spin will negate 6 plaquette terms.  Therefore, the first order correction to the Wilson surface is
\begin{align}
    \left \langle \prod_{i \in \partial V} s_i \right \rangle = \frac{\sum_{\vec{s}} \left(\prod_{i \in \partial V} s_i\right) e^{-\beta H(\vec{s})}}{\sum_{\vec{s}} e^{-\beta H(\vec{s})}} = \frac{1 + (N - 2L)e^{-16 \beta} + N e^{-12 \beta} + \cdots}{1 + N e^{-16 \beta} + N e^{-12 \beta} + \cdots}
\end{align}
where $L$ is the length of the loop.  For $n$ spin flips, we make the assumption that they are independent.  Then, the contribution $F(n)$ to the numerator is
\begin{equation}
  F(n) = \sum_{a+b = n} \frac{(N-2L)^a}{a\,!} \frac{N^b}{b\,!}  e^{-\beta (16a + 12b)} 
\end{equation}
and the contribution to the denominator $G(n)$ is
\begin{equation}
  G(n) = \sum_{a+b = n} \frac{N^a}{a\,!} \frac{N^b}{b\,!}  e^{-\beta (16a + 12b)}.
\end{equation}
Then, the Wilson surface roughly has the expectation value
\begin{align}
    \left \langle \prod_{i \in \partial V} s_i \right \rangle &\approx \frac{1 + F(1) + F(2) +  \cdots}{1 + G(1) + G(2) +  \cdots} \nonumber \\
    &= \frac{\exp\left((N-2L)e^{-16\beta} + N e^{-12\beta}\right)}{\exp\left(N e^{-16\beta} + N e^{-12\beta}\right)} \nonumber \\
    &= \exp\left(-2L\cdot e^{-16\beta}\right),
\end{align}
which is representative of the deconfined phase.

Next, we understand what happens in the confined phase.  For this, we need to perform a high temperature expansion of the partition function:
\begin{equation}
    \mathcal{Z} = \left(\cosh \beta\right)^N \sum_{\vec{s}} \prod_p (1 + \tanh \beta \, (s_i s_j s_k s_\ell \tau_m)).
\end{equation}
and the Wilson loop looks like
\begin{equation}
     \left \langle \prod_{i \in C} s_i \right \rangle = \frac {\sum_{\vec{s}} \left(\prod_{i \in \partial V} s_i\right) \prod_p (1 + \tanh \beta \, (s_i s_j s_k s_\ell \tau_m))}{\sum_{\vec{s}} \prod_p (1 + \tanh \beta \, (s_i s_j s_k s_\ell \tau_m))}.
\end{equation}
Naively, in analogy to the Ising gauge theory, one may fill the interior of $C$ with plaquettes.  Doing this results in $A$ dangling ghost spins, where $A$ is the area of the loop.  These ghost spins are sites on a 2D square lattice with boundary $C$.  Next, we may rewrite a correlation function of two ghost spins as
\begin{equation}
    \tau_i \tau_j = \prod_{k = i}^{j} \tau_k \left(\prod_{\alpha \in \square_k} s_{\alpha}\right)^2 \tau_{k+1}
\end{equation}
where the first product $\prod_{k = i}^{j}$ is over a string connecting $i$ and $j$ and the second product $\prod_{\alpha \in \square_k} s_{\alpha}$ is over bond spins on the plaquette adjacent to ghost spin $\tau_k$ and pointing out of the plane of ghost spins.  This identity may be written as a product of a string of pyramidal operators connecting $\tau_i$ and $\tau_j$.  To satisfy each ghost spin in the 2D substructure, we need to connect pairs of them together via strings of pyramidal operators while minimizing the total number of pyramidal operators.  The way to do this is for the string operators to mimic dimer configurations on the square lattice.  There are 2 pyramids involved per dimer, and therefore $A$ pyramids involved in a given dimer configuration.  Then, the leading order contribution to the Wilson operator looks like
\begin{equation}
    \left \langle \prod_{i \in C} s_i \right \rangle \sim \left(\tanh \beta\right)^{A}\cdot D_S \left(\tanh \beta\right)^A + \cdots,
\end{equation}
where $D_S$ is the number of dimer configurations on the square lattice with $A$ sites, 
\begin{equation}
    \lim_{A \to \infty} \frac{1}{A} \log D_S = \frac{2 G}{\pi},
\end{equation}
and $G$ is Catalan's constant.  Therefore, the Wilson surface obeys an area law in the confined phase for large enough temperatures. 

If the Wilson loop encloses an odd number of plaquettes, then filling the interior or any extrusion of the loop with plaquettes will result in an odd number of dangling ghost spins.  These cannot be paired together without excluding a single ghost spin.  This single ghost spin can only be resolved by a string of pyramids connecting the ghost spin to the boundary of the lattice.  The Wilson loop then behaves like
\begin{equation}
    \left \langle \prod_{i \in C} s_i \right \rangle  \lesssim D_S \left(\tanh^2 \beta\right)^{A}\cdot (6 \tanh\beta)^{\sqrt[3]{N}}  + \cdots,
\end{equation}
which for small enough $\beta$ vanishes in the thermodynamic limit $N \to \infty$.

\section{Simulation details}

\begin{figure}
    \centering
    \includegraphics[scale=0.4]{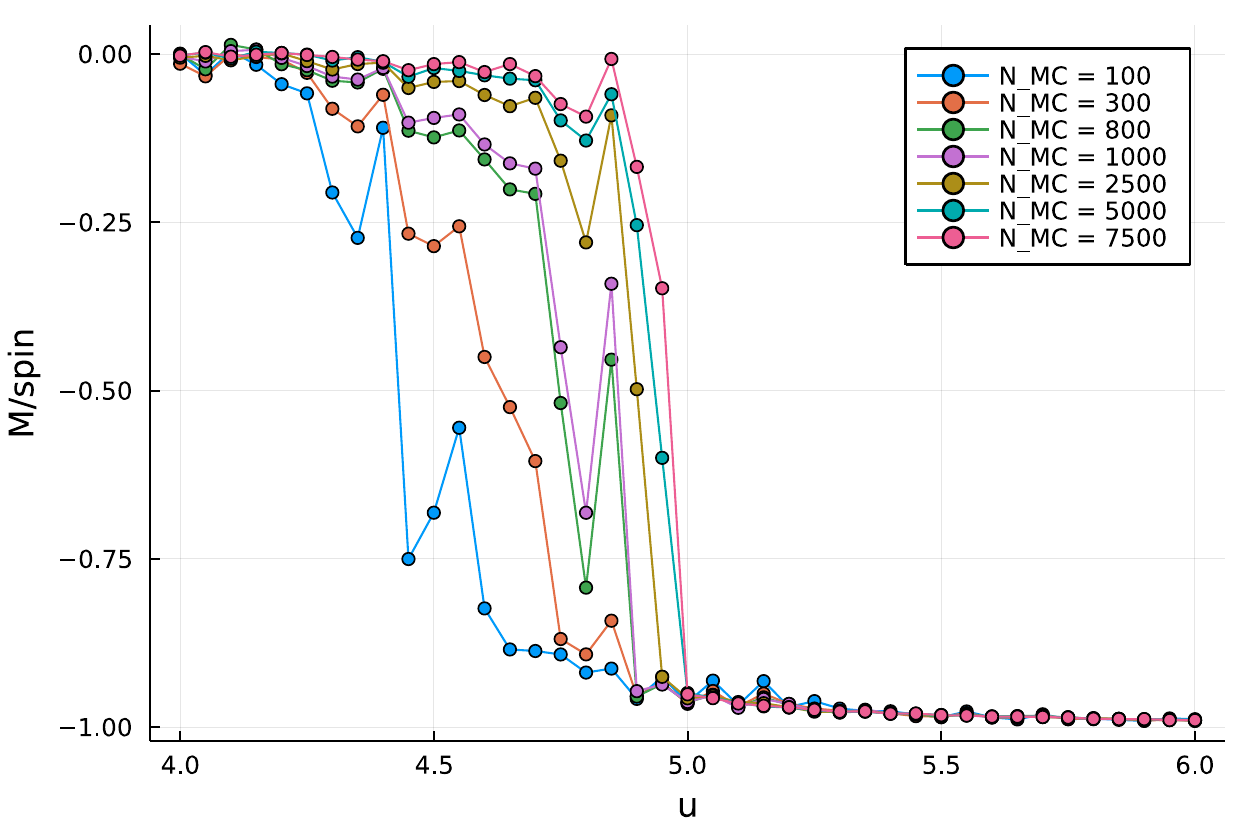}
    \caption{The Monte Carlo results show that as the number of samples increases, the transition point converges to the self dual point at $u = 5$.  The magnetization per spin is plotted.  Similar plots were generated for the heat capacity, which indicates that there is a single unique transition.}
    \label{fig:suppmag}
\end{figure}

The Monte Carlo simulation consists of loop moves implemented as Metropolis updates.  While the plot shown in the main text provides convincing evidence for a transition at $u = 5$ for the octahedral lattice, here we provide further plots to corroborate this.  In particular, we plot the magnetization as a function of the number of samples (which is proportional to the total number of updates).  We see that the transition point indeed robustly converges to $u = 5$.  The results are shown in Fig.~\ref{fig:suppmag}.  Similar results were seen for the diamond and BCC lattices upon finite size scaling.

\section{Simple proof of spontaneously magnetized phase}

We present a short and rigorous proof that the symmetry enriched $\mathbb{Z}_2$
 gauge theory for the cubic lattice model exhibits a phase transition in the local order parameter, and will thus arrive at a bound on the transition temperature (even though in the main text, the transition temperature is computed exactly).  To do this, we will rely on the Griffiths-Kelly-Sherman inequalities, which state that for a classical spin model with purely ferromagnetic multispin interactions with coupling constants $J_1, J_2, \cdots$, the expectation value of a product of spin operators, $\mathcal{O} = \prod_{i \in S} s_i$ where $S$ is some set, satisfies
 \begin{equation}
     \frac{\partial \langle \mathcal{O}(J_1, J_2, \cdots) \rangle}{\partial J_i} \geq 0.
 \end{equation}
 Note that the current 5-spin interaction model has an antiferromagnetic coupling.  However, we can make the transformation $\tau \to -\tau$ to the ghost spins without changing the partition function, rendering the interaction ferromagnetic and applicable to the inequality.  In particular, for the 5-spin interaction model, we remove pyramidal interactions everywhere except for a two-dimensional layer of cubes.  We then remove the pyramids on the top and the bottom of each cube, and as a result, we know that the magnetization can only decrease as a result of removing all of these interactions.  The corresponding classical spin Hamiltonian will then take the form
 \begin{equation}
     \mathcal{H} = J \sum_{p'} s_1 s_2 s_3 s_4 \tau_5
 \end{equation}
 where the $'$ symbol indicates the sum is only over the remaining pyramids.  Among the 4 bond spins $s_1$, $s_2$, $s_3$, and $s_4$, two spins are shared by \emph{two} pyramids, while the other two spins are shared by \emph{four} pyramids.  If we call the former spins $\overline{s}$, then
 \begin{equation}
     \sum_{\overline{s}} \exp\left(J \sum_{p'} \overline{s}_1 \overline{s}_2 s_3 s_4 \tau_5\right) \propto \prod_{\langle p,q \rangle} \cosh\left(J s_3 s_4
     (\tau_p + \tau_q)\right)
 \end{equation}
 where $\langle p,q \rangle$ correspond to edges in the square lattice formed by the sites of the ghost spins.  Through the standard trick
 \begin{equation}
     \sum_{s_3, s_4} \cosh\left(J s_3 s_4
     (\tau_p + \tau_q)\right) = 2 \cosh\left(J
     (\tau_p + \tau_q)\right)
 \end{equation}
 and the identity 
 \begin{equation}
2 \cosh\left(J
     (\tau_p + \tau_q)\right) = 1 - \tau_p \tau_q + (1 + \tau_p \tau_q)\cosh 2J
 \end{equation}
 The partition function becomes
 \begin{equation}
     \mathcal{Z} \propto \sum_{\tau} \exp \left(J' \sum_{\langle p, q \rangle} \tau_p \tau_q\right),
 \end{equation}
 where
 \begin{equation}
     \tanh J' = \frac{\cosh 2J - 1}{\cosh 2J + 1}.
 \end{equation}
 In 2D, we know that the Ising model spontaneously magnetizes, and since the magnetization in the 5-spin model has a strictly greater magnetization, it must magnetize as well.  This proves the existence of a symmetry broken phase.  The critical temperature of the 2D Ising model on a square lattice satisfies $\tanh J_{\text{ising}} = \sqrt{2}-1$, and thus we find that
 \begin{equation}
     u = \frac{\cosh (6J)}{\cosh (2J)} = 9 + 8\sqrt{2} > 5
 \end{equation}
 is an upper bound on the value of $u$ for a phase transition, since for greater values of $u$ we have proven that the 5-spin interaction Hamiltonian must spontaneously magnetize.  The true phase transition occurs (of course) at $u = 5$.

\section{Explicit construction of Hamiltonian in terms of spin operators}

In this appendix, we provide an explicit form for the parent Hamiltonian in the main text.  

To construct a parent Hamiltonian whose ground state is the RK vertex model wavefunction, place spins on the edges of the diamond/cubic/BCC lattices (alternatively on the sites of the dual lattices) with the interpretation that spin up corresponds to the presence of a dimer and spin down the absence of a dimer.  Define $P = -\sum_v \prod_{i \in v} Z_i$ where $v$ is a site on the lattice and $i \in v$ is the set of edges touching $v$; this operator projects onto the set of configurations obeying an even dimer constraint.  Call $\ell$ a loop of minimal length (squares on the cubic and BCC lattices and hexagons on the diamond lattice) and for a configuration $C$ of dimers, define $W_\ell(C_\ell) = \prod_{i \in \ell} W_i(C_\ell)$, where $i$ are sites along $\ell$ and $W_i$ denotes the vertex weight of the dimer configuration on $i$.  Furthermore, define $W_\ell(\overline{C}_\ell)$ to be the product of vertex weights if all the spins on $\ell$ are flipped.  Define $\omega(C_\ell, \overline{C}_\ell) = \sqrt{W_\ell(C_\ell)/W_\ell(\overline{C}_\ell)}$.  Then, the local frustration-free Hamiltonian $H = P + H_A + r H_B$, where
\begin{align}
    H_A &= \sum_{C_{\ell}, \ell} \frac{1}{\omega(C_\ell, \overline{C}_\ell)} \ketbra{C_{\ell}}{C_{\ell}} + \omega(C_\ell, \overline{C}_\ell) \ketbra{\overline{C}_{\ell}}{\overline{C}_{\ell}} \label{eq:fullham}\\
    H_B &= -\sum_{C_{\ell}, \ell} \ketbra{C_{\ell}}{\overline{C}_{\ell}} + \ketbra{\overline{C}_{\ell}}{C_{\ell}} \label{eq:fullham2}
\end{align}
has the desired ground state at $r = 1$.  This follows because at $r=1$, the Hamiltonian $H_A + H_B$ gives
\begin{equation}
    H_A + H_B = \sum_{C_{\ell}, \ell} \left(\frac{1}{\sqrt{\omega(C_\ell, \overline{C}_\ell)}} \ket{C_{\ell}} - \sqrt{\omega(C_\ell, \overline{C}_\ell)} \ket{\overline{C}_{\ell}}\right)\left(\frac{1}{\sqrt{\omega(C_\ell, \overline{C}_\ell)}} \bra{C_{\ell}} - \sqrt{\omega(C_\ell, \overline{C}_\ell)} \bra{\overline{C}_{\ell}}\right).
\end{equation}

\begin{figure*}
    \centering
    \includegraphics[scale=0.5]{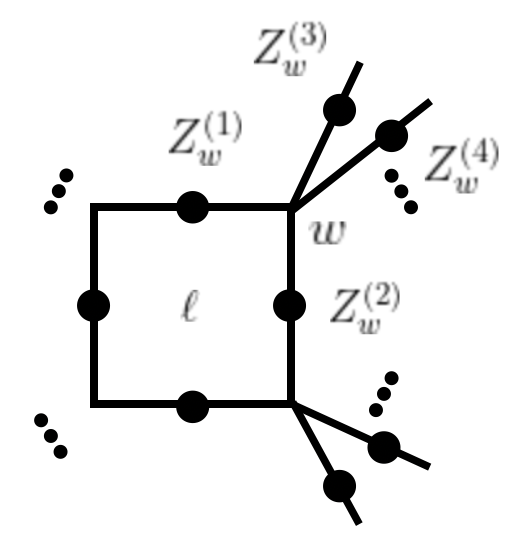}
    \caption{The location of the spins denoted by $Z_w^{(i)}$}
    \label{fig:spinloc}
\end{figure*}

The RK wavefunction is annihilated by each of these projectors, and is thus a ground state.  We now will write down explicit expressions for the weights $\omega(C_\ell, \overline{C}_\ell)$.  Restricting to the cubic and diamond lattices, given a configuration of dimers surrounding a vertex, the Boltzmann weight is
\begin{equation}
W(Z_1, Z_2, \cdots, Z_V) = e^{J_V (Z_1 + Z_2 + \cdots + Z_V)^2}
\end{equation}
where $V = 4,6$, while $e^{16 J_4} = u$ and $e^{32 J_6} = v$.  For the BCC lattice,
\begin{equation}
W(Z_1, Z_2, \cdots, Z_8) = e^{J_{1,8} (Z_1 + Z_2 + \cdots + Z_8)^2 + J_{2,8} (Z_1 + Z_2 + \cdots + Z_8)^4}
\end{equation}
and $J_{1,8}$ and $J_{2,8}$ are given in Eqn.~\ref{eqn:couplings}.  As a consequence, we may write the weights $\omega(C_\ell, \overline{C}_\ell)$ for the diamond and cubic lattices as
\begin{align}
\omega(C_\ell, \overline{C}_\ell) &= \exp(\frac{J_V}{2}\sum_{w\in \ell} (Z_w^{(1)} + Z_w^{(2)} + \cdots + Z_w^{(V)})^2 - (-Z_w^{(1)} - Z_w^{(2)} + \cdots + Z_w^{(V)})^2) \\
&= \exp(2 J_V \sum_{w\in \ell} (Z_w^{(1)} + Z_w^{(2)})(Z_w^{(3)} + \cdots + Z_w^{(V)})).
\end{align}
where $w$ indicates a site along loop $\ell$ and $Z_w^{(1)}, Z_w^{(2)}$ are the two spins on links that touch site $u$ but are on the loop $\ell$.  The remaining $Z_w^{(3)}, Z_w^{(4)},\cdots$ are spins on links that touch $u$ but do not reside on $\ell$.  A visual for the location of these spins is shown in Figure~\ref{fig:spinloc} (for simplicity $\ell$ is shown as a square plaquette).
For the BCC lattice, there is an additional term and the weights are given by
\begin{align}\label{eqn:bccweights}
\omega(C_\ell, \overline{C}_\ell) = \exp\left(\sum_{w\in \ell} 2 J_{1,8} (Z_w^{(1)} + Z_w^{(2)})(Z_w^{(3)} + \cdots + Z_w^{(V)}) + 4 J_{2,8} (Z_w^{(1)} + Z_w^{(2)})^3(Z_w^{(3)} + \cdots + Z_w^{(V)}) \right. \nonumber \\ \left. + 4 J_{2,8} (Z_w^{(1)} + Z_w^{(2)})(Z_w^{(3)} + \cdots + Z_w^{(V)})^3\right).
\end{align}
As such, we may write the Hamiltonian (for the diamond and cubic lattices)
\begin{equation}
 H_A = \sum_{C_{\ell}, \ell} \frac{1}{\omega(C_\ell, \overline{C}_\ell)} \ketbra{C_{\ell}}{C_{\ell}} + \omega(C_\ell, \overline{C}_\ell) \ketbra{\overline{C}_{\ell}}{\overline{C}_{\ell}} = \sum_{\ell} \exp(-2 J_V \sum_{w\in \ell} (Z_w^{(1)} + Z_w^{(2)})(Z_w^{(3)} + \cdots + Z_w^{(V)}))
\end{equation}
and a similar Hamiltonian can be written for the BCC lattice.  Next, we note that the kinetic term responsible for ring exchanges is given by
\begin{equation}
H_B = -\sum_{C_{\ell}, \ell} \ketbra{C_{\ell}}{\overline{C}_{\ell}} + \ketbra{\overline{C}_{\ell}}{C_{\ell}} = -\sum_{\ell} \prod_{\alpha \in \ell} X_{\alpha},
\end{equation}
where $\alpha \in \ell$ label links on loop $\ell$.  In this appendix, lower case Latin letters denote sites, and lower case Greek letters denote links (on which the dimers live).  Therefore, the full Hamiltonian (for the cubic and diamond lattice vertex models) is given by
\begin{equation}
H = \sum_{\ell} \exp(-2 J_V \sum_{w\in \ell} (Z_w^{(1)} + Z_w^{(2)})(Z_w^{(3)} + \cdots + Z_w^{(V)})) - \sum_{w} \prod_{\alpha \in w} Z_{\alpha}  - \sum_{\ell} \prod_{\alpha \in \ell} X_{\alpha}.
\end{equation}
where $\alpha \in w$ label the set of links touching site $w$.  We note that this Hamiltonian commutes with the $\mathbb{Z}_2$ gauge constraint $\prod_{\alpha \in w} Z_{\alpha}$ as well as the global $\mathbb{Z}_2$ symmetry $\prod_{\alpha} X_{\alpha}$. Therefore, one should think of this theory as a $\mathbb{Z}_2$ gauge theory with an additional global symmetry, and not as a gauge theory coupled to matter fields.  We also recognize that when $u = 1$, $J_V = 0$, and the theory reduces to the 3D toric code.  Expanding out the exponential, one can see that the effect of the first term is to add additional multi-spin interactions to the traditional toric code Hamiltonian, whilst preserving the global $\mathbb{Z}_2$ symmetry.  A similar Hamiltonian can be written down for the BCC lattice quantum vertex model using Eqn.~\ref{eqn:bccweights}.

\section{Arrowed quantum vertex models}

In this appendix, we derive an alternative formulation of the quantum vertex model where the dimer variables are replaced with arrow variables.  Because both the cubic and diamond lattices are \emph{bipartite} and the vertex weights possess a $\mathbb{Z}_2$ symmetry, any configuration of dimers can be mapped onto a configuration of arrows, where a dimer corresponds to an arrow pointing out of the A sublattice and pointing into the B sublattice.  While for bipartite lattices arrow and dimer variables are equivalent, on nonbipartite lattices, these constraints yield vastly different phase diagrams.  For a bond vertex model on a non-bipartite lattice, we expect to find a rather generic gapless point, while for an arrowed vertex model on the same lattice, we expect the phase diagram to host a single $\mathbb{Z}_2$ topologically ordered phase.  This has been established numerically for a couple of examples in Ref.~\cite{Balasubramanian} in 2D, and we expect it to hold true in higher dimensions.  For this reason, it is worth discussing the construction of arrowed vertex models. 

\begin{figure*}
    \centering
    \includegraphics[scale=0.5]{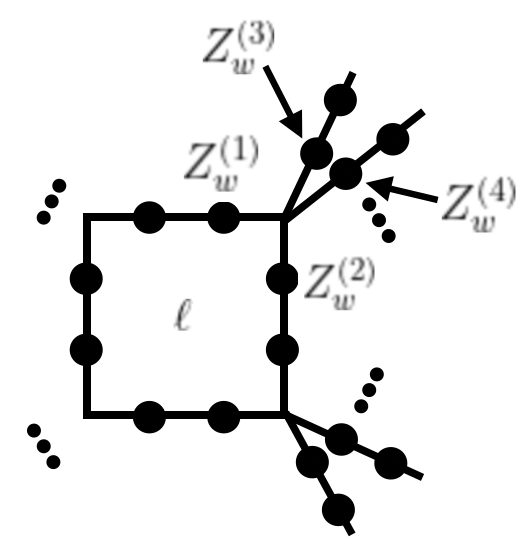}
    \caption{The location of the spins denoted by $Z_w^{(i)}$ for the arrowed vertex model.  An antiferromagnetic interaction occurs for each pair of spins on a bond}
    \label{fig:spinloc2}
\end{figure*}

The construction of the arrowed vertex models requires a small modification: on each edge of the dual lattice (for example, a cubic or diamond lattice) we place \emph{two} spins rather than one, and we project onto configurations where the spins are antiparallel to each other by adding the term $Z_i Z_j$ for each pair of spins $\langle i, j \rangle$ on an edge.  Antiparallel configurations of these spins are in one-to-one correspondence with arrows (as the up spin corresponds to the arrow head and the down spin corresponds to the tail).  The Hamiltonian is then equivalent to the constructions in the previous appendix with some minor changes.  First, one adds the antiferromagnetic term $\sum_{\langle i,j \rangle} Z_i Z_j$ to the Hamiltonian.  Second, the exchange term $\prod_{\alpha \in \ell} X_{\alpha}$ is changed to $\prod_{\langle \alpha, \beta \rangle \in \ell} X_{\alpha} X_{\beta}$.  Finally, when constructing the Boltzmann weight $\omega(C_{\ell}, \overline{C}_{\ell})$, the spins which are used are indicated in Figure~\ref{fig:spinloc2}

For bipartite lattices, such as the diamond or cubic lattices, the phase diagram and transitions are identical to that of the quantum vertex model with dimer variables as opposed to arrow variables.  On non-bipartite lattices, the story is different, and the topologically ordered phase is more robust.  This follows from that fact that the $u = \infty$ point no longer exhibits spontaneous symmetry breaking and can have extensive ground state degeneracy.  In two dimensions, for a lattice of corner sharing triangles, we have previously shown that the arrowed vertex model with $W_4(0) = W_4(4) = u$ and $W_4(2) = 1$ has ground state configurations which are in one-to-one correspondence with three coloring configurations on that lattice when $u \to \infty$.  The quantum vertex model becomes a quantum three-coloring model in this limit, which is gapless.  A similar mapping to a three coloring model occurs for the hyperkagome lattice in 3D.  For all other values of $u$, we expect this model to be in a topologically ordered phase.  This is evidenced by the fact that $u = 1$ corresponds to a toric code-like wavefunction exhibiting $\mathbb{Z}_2$ topological order, even for the arrowed vertex model.

\end{document}